\documentclass[aps,twocolumn,showkeys,showpacs,preprintnumbers,prd,superscriptaddress,nofootinbib,10pt]{revtex4-1}
\usepackage{graphicx,epsf,bm,amsmath,amsfonts,amssymb,epstopdf,natbib,hyperref,color,verbatim,multirow,bm}
\hypersetup{colorlinks=true,urlcolor=blue,citecolor=blue,linkcolor=blue,menucolor=blue,anchorcolor=blue,filecolor=blue}
\usepackage{appendix}
\date{\today}

\begin{document}

\title{Primordial regular black holes as all the dark matter. \\ I. Time-radial-symmetric metrics}

\author{Marco Calz\`{a}}
\email{marco.calza@unitn.it}
\affiliation{Department of Physics, University of Trento, Via Sommarive 14, 38123 Povo (TN), Italy}
\thanks{M.C. and D.P. contributed equally to this work}
\affiliation{Trento Institute for Fundamental Physics and Applications (TIFPA)-INFN, Via Sommarive 14, 38123 Povo (TN), Italy}

\author{Davide Pedrotti}
\email{davide.pedrotti-1@unitn.it}
\affiliation{Department of Physics, University of Trento, Via Sommarive 14, 38123 Povo (TN), Italy}
\thanks{M.C. and D.P. contributed equally to this work}
\affiliation{Trento Institute for Fundamental Physics and Applications (TIFPA)-INFN, Via Sommarive 14, 38123 Povo (TN), Italy}

\author{Sunny Vagnozzi}
\email{sunny.vagnozzi@unitn.it}
\affiliation{Department of Physics, University of Trento, Via Sommarive 14, 38123 Povo (TN), Italy}
\affiliation{Trento Institute for Fundamental Physics and Applications (TIFPA)-INFN, Via Sommarive 14, 38123 Povo (TN), Italy}

\begin{abstract}
\noindent Primordial black holes (PBHs) are usually assumed to be described by the Schwarzschild or Kerr metrics, which however feature unwelcome singularities. We study the possibility that PBHs are non-singular objects, considering three phenomenological, regular \textit{tr} (time-radial)-symmetric space-times (including the well-known Bardeen and Hayward ones), featuring either de Sitter or Minkowski cores. We characterize the evaporation of these PBHs and constrain their abundance from $\gamma$-ray observations. For all three metrics we find that constraints on $f_{\text{pbh}}$, the fraction of dark matter (DM) in the form of PBHs, weaken with respect to the Schwarzschild limits, because of modifications to the PBH temperature and greybody factors. This moves the lower edge of the asteroid mass window down by potentially an order of magnitude or more, leading to a much larger region of parameter space where PBHs can make up all the DM. A companion paper is devoted to non-\textit{tr}-symmetric metrics, including loop quantum gravity-inspired ones. Our work provides a proof-of-principle for the interface between the DM and singularity problems being a promising arena with a rich phenomenology.
\end{abstract}

\maketitle

\section{Introduction}
\label{sec:introduction}

The Standard Model of particle physics (SM) and General Relativity (GR) have proven to be extremely successful at describing a huge range of terrestrial, astrophysical, and cosmological observations. However, their successes are limited by a number of shortcomings, potentially (especially in the case of SM) pointing towards the need for new physics which may better describe the matter and gravity sectors. On the more observational/phenomenological side, the SM lacks a candidate for the dark matter (DM) which accounts for $\simeq 25\%$ of the energy budget of the Universe~\cite{Arbey:2021gdg,Cirelli:2024ssz}. On the more theoretical side, continuous gravitational collapse in GR leads to the pathological appearance of curvature singularities~\cite{Penrose:1964wq,Hawking:1970zqf}. The nature of DM and the singularity problem are arguably two among the most important open questions in theoretical physics.

The solution to the DM problem could reside in the physics of some of the most peculiar objects in the Universe: black holes (BHs). It has long been realized that primordial BHs (PBHs), hypothetical relics from the primordial Universe formed from the collapse of large density perturbations upon horizon re-entry, are indeed excellent DM candidates~\cite{Chapline:1975ojl,Meszaros:1975ef,Khlopov:1980mg,Khlopov:1985fch,Ivanov:1994pa,Choudhury:2013woa,Belotsky:2014kca,Bird:2016dcv,Clesse:2016vqa,Poulin:2017bwe,Raccanelli:2017xee,LuisBernal:2017fmf,Clesse:2017bsw,Kohri:2018qtx,Liu:2018ess,Liu:2019rnx,Murgia:2019duy,Carr:2019kxo,Liu:2020cds,Hertzberg:2020hsz,Serpico:2020ehh,DeLuca:2020bjf,DeLuca:2020fpg,DeLuca:2020qqa,Carr:2020erq,Bhagwat:2020bzh,DeLuca:2020sae,Wong:2020yig,Carr:2020mqm,Domenech:2020ssp,DeLuca:2021wjr,Arbey:2021ysg,Franciolini:2021tla,DeLuca:2021hde,Cheek:2021odj,Cheek:2021cfe,Heydari:2021gea,Dvali:2021byy,Heydari:2021qsr,DeLuca:2021pls,Liu:2021jnw,Saha:2021pqf,Bhaumik:2022pil,Anchordoqui:2022txe,Cai:2022erk,Oguri:2022fir,Franciolini:2022tfm,Mazde:2022sdx,Cai:2022kbp,Anchordoqui:2022tgp,Liu:2022iuf,Fu:2022ypp,Choudhury:2023vuj,Papanikolaou:2023crz,deFreitasPacheco:2023hpb,Choudhury:2023jlt,Choudhury:2023rks,Musco:2023dak,Yuan:2023bvh,Choudhury:2023hvf,Ghoshal:2023sfa,Cai:2023uhc,Choudhury:2023kdb,Huang:2023chx,Choudhury:2023hfm,Bhattacharya:2023ysp,Heydari:2023xts,Heydari:2023rmq,Choudhury:2023fwk,Choudhury:2023fjs,Ghoshal:2023pcx,Hai-LongHuang:2023atg,Huang:2023mwy,Anchordoqui:2024akj,Choudhury:2024one,Thoss:2024hsr,Papanikolaou:2024kjb,Choudhury:2024ybk,Choudhury:2024jlz,Anchordoqui:2024dxu,Papanikolaou:2024fzf,Yin:2024xov,Choudhury:2024dei,Heydari:2024bxj,Dvali:2024hsb,Boccia:2024nly,Hai-LongHuang:2024kye,Choudhury:2024dzw,Anchordoqui:2024jkn,Yang:2024vij,Saha:2024ies,Anchordoqui:2024tdj,Chen:2024pge,Dai:2024guo,Hai-LongHuang:2024vvz,Zantedeschi:2024ram,Chianese:2024rsn,Barker:2024mpz,Borah:2024bcr,Hai-LongHuang:2024gtx} (see e.g.\ Refs.~\cite{Khlopov:2008qy,Carr:2016drx,Green:2020jor,Carr:2020xqk,Villanueva-Domingo:2021spv,Carr:2021bzv,Bird:2022wvk,Carr:2023tpt,Arbey:2024ujg,Choudhury:2024aji} for reviews): in fact, PBHs are the only viable DM candidate which does not invoke new particles surviving to the present day. Once believed to merely be objects of mathematical speculation, observational effects associated to BHs are now routinely detected~\cite{Bambi:2017iyh}, turning these objects into extraordinary probes of fundamental physics~\cite{Creminelli:2017sry,Sakstein:2017xjx,Ezquiaga:2017ekz,Boran:2017rdn,Baker:2017hug,Amendola:2017orw,Visinelli:2017bny,Crisostomi:2017lbg,Dima:2017pwp,Cai:2018rzd,Casalino:2018tcd,Barack:2018yly,LIGOScientific:2018dkp,Casalino:2018wnc,Held:2019xde,Bambi:2019tjh,Vagnozzi:2019apd,Zhu:2019ura,Cunha:2019ikd,Banerjee:2019nnj,Banerjee:2019xds,Allahyari:2019jqz,Vagnozzi:2020quf,Khodadi:2020jij,Kumar:2020yem,Khodadi:2020gns,Pantig:2021zqe,Khodadi:2021gbc,Roy:2021uye,Uniyal:2022vdu,Pantig:2022ely,Ghosh:2022kit,Khodadi:2022pqh,KumarWalia:2022aop,Shaikh:2022ivr,Odintsov:2022umu,Oikonomou:2022tjm,Pantig:2023yer,Gonzalez:2023rsd,Sahoo:2023czj,Nozari:2023flq,Uniyal:2023ahv,Filho:2023ycx,Raza:2023vkn,Hoshimov:2023tlz,Chakhchi:2024tzo,Liu:2024lbi,Liu:2024lve,Khodadi:2024ubi,Liu:2024soc,Nojiri:2024txy}. As a result, the possibility of PBHs accounting for the entire DM budget is severely constrained by a wide range of considerations and constraints. The only (not entirely undebated) remaining open window of parameter space where PBHs could make up all the DM is the so-called ``\textit{asteroid mass window}'', roughly for PBH masses $10^{17}\,{\text{g}} \lesssim M_{\text{pbh}} \lesssim 10^{23}\,{\text{g}}$~\cite{Katz:2018zrn,Bai:2018bej,Smyth:2019whb,Coogan:2020tuf,Ray:2021mxu,Auffinger:2022dic,Ghosh:2022okj,Miller:2021knj,Branco:2023frw,Bertrand:2023zkl,Tran:2023jci,Gorton:2024cdm,Dent:2024yje,Tamta:2024pow,Tinyakov:2024mcy,Loeb:2024tcc}: lighter PBHs would have evaporated fast enough to either have disappeared by now or overproduced $\gamma$-rays in the ${\text{MeV}}$ range, whereas heavier PBHs would have been detected through the microlensing of background stars.

Almost all works on PBHs assume that these are Schwarzschild or Kerr BHs~\cite{Khlopov:2008qy,Carr:2016drx,Green:2020jor,Carr:2020xqk,Villanueva-Domingo:2021spv,Carr:2021bzv,Bird:2022wvk,Carr:2023tpt,Arbey:2024ujg,Choudhury:2024aji}. All constraints and considerations on DM potentially being in the form of PBHs are therefore subject to this underlying assumption. The existence of the asteroid mass window, and the extension thereof, is of course no exception. The assumption in question is not at all unreasonable at least from the phenomenological point of view, given that there are at present no signs of tension between astrophysical observations and the Kerr-Newman family of metrics, and more generally the no-hair theorem. Nevertheless, from the theoretical point of view such an assumption might stir some unease, given the appearance of singularities in the Schwarzschild and Kerr metrics. The above considerations naturally lead to the following question: ``\textit{what if PBHs are non-singular}''? The present work is a pilot study whose goal is to explore this question, which naturally merges the DM and singularity problems.

We entertain the possibility that PBHs are ``regular'', i.e.\ free of curvature singularities~\cite{Ansoldi:2008jw,Nicolini:2008aj,Sebastiani:2022wbz,Torres:2022twv,Lan:2023cvz}, and therefore that DM may be in the form of primordial regular BHs (PRBHs). For concreteness, we consider so-called \textit{tr} (time-radial)-symmetric metrics, for which the product of the coefficients for the $dt^2$ and $dr^2$ components of the line element in four-dimensional Boyer–Lindquist coordinates is equal to $-1$, and the function which multiplies the angular part of the line element is $r^2$, i.e.\ $r$ is the areal radius. More specifically, we focus on the following three regular, static spherically symmetric space-times, all of which are characterized by an additional \textit{regularizing} parameter $\ell$ and recover the Schwarzschild space-time in the $\ell \to 0$ limit: Bardeen BHs~\cite{Bardeen:1968ghw}, Hayward BHs~\cite{Hayward:2005gi}, and Culetu-Ghosh-Simpson-Visser BHs~\cite{Culetu:2013fsa,Culetu:2014lca,Ghosh:2014pba,Simpson:2019mud}. These three space-times present a rich phenomenology, including either de Sitter or Minkowski cores. We focus our attention on observational constraints from PRBH evaporation, which set the lower edge of the asteroid mass window, discussing in detail how the evaporation process is modified with respect to that for Schwarzschild PBHs. We show that, as a result, the phenomenology of PRBHs can be quite different from that of Schwarzschild PBHs, with a larger range of masses where PRBHs could make up the entire DM component, opening up the asteroid mass window by up to an extra decade in mass. Keeping in mind that the metrics in question are phenomenological in nature, our results demonstrate that a common solution to the DM and singularity problems in the form of PRBHs is one which is worth taking seriously, and warrants further investigation, and more generally the interface of these two problems provides a promising arena. We stress that our work should not be intended as a comprehensive analysis of PRBHs, but rather as a pilot study, pointing towards a direction which has thus far received little attention and indicating promising directions for further work.

The rest of this paper is then organized as follows. In Sec.~\ref{sec:regular} we briefly introduce the regular space-times studied in the rest of the work. Various aspects of our methodology are discussed in Sec.~\ref{sec:methodology}, with Sec.~\ref{subsec:greybody} devoted to the calculation of the so-called greybody factors, Sec.~\ref{subsec:spectra} to the computation of photon spectra resulting from Hawking evaporation, and Sec.~\ref{subsec:constraints} to the comparison against observations. The resulting limits on the fraction of DM which may be in the form of primordial regular BHs are then critically discussed in Sec.~\ref{sec:results}. Finally, in Sec.~\ref{sec:conclusions} we draw concluding remarks. A number of more technical aspects concerning the greybody factors computation are discussed in Appendix~\ref{sec:appendix}, whereas the time evolution of the primordial regular BHs we consider is studied in Appendix~\ref{sec:appendixb}. Unless otherwise specified, we adopt units where $G=c=\hbar=1$. In closing, we note that a related study is being presented in a companion paper~\cite{Calza:2024xdh}: this focuses on non-\textit{tr}-symmetric metrics, which also include loop quantum gravity-inspired metrics, but at the same time complicate the study of the evaporation process. We recommend that the interested reader go through the present work prior to consulting our companion paper~\cite{Calza:2024xdh}.

\section{Regular black holes}
\label{sec:regular}

It is well known, thanks to the Penrose-Hawking singularity theorems, that continuous gravitational collapse in GR sourced by matter contents satisfying reasonable energy conditions leads to the appearance of singularities~\cite{Penrose:1964wq,Hawking:1970zqf}. These are regions of space-time where curvature invariants, i.e.\ sets of independent scalars constructed from the Riemann tensor and the metric, diverge (with the archetypal example being the central singularity in the Kerr-Newman family of metrics). These singularities are arguably unsatisfactory as they lead to a potential breakdown in predictivity (see Refs.~\cite{Sachs:2021mcu,Ashtekar:2021dab,Ashtekar:2022oyq} for a different viewpoint). For this reason, they are oftentimes regarded as a manifestation of our lack of knowledge of (new) physics in the high-energy/high-curvature regime. A widespread belief is that quantum gravity effects on these scales would ultimately cure the singularity problem (and potentially lead to observable effects), although this is more of a hope supported only by a few first-principles studies~\cite{Dymnikova:1992ux,Dymnikova:2004qg,Ashtekar:2005cj,Bebronne:2009mz,Modesto:2010uh,Spallucci:2011rn,Perez:2014xca,Colleaux:2017ibe,Nicolini:2019irw,Bosma:2019aiu,Jusufi:2022cfw,Olmo:2022cui,Jusufi:2022rbt,Ashtekar:2023cod,Nicolini:2023hub}.

Even in the absence of a widely agreed upon theory of quantum gravity, one can still hope to make progress in understanding and taming singularities, while also potentially gaining intuition about the possible features of such a theory, through a more phenomenological approach. Under the assumption that a metric description maintains its validity, one can introduce metrics which are free of singularities in the entire space-time, and describe so-called regular BHs (RBHs)~\cite{Bambi:2023try}. It is often (albeit not necessarily always) the case that RBH metrics are controlled by an extra parameter, which in what follows we shall refer to as \textit{regularizing parameter} (and denote by $\ell$), typically recovering the Schwarzschild metric (for non-rotating RBHs) in the limit $\ell \to 0$. Several RBH metrics have been studied over the past decades, see e.g.\ Refs.~\cite{Borde:1996df,AyonBeato:1998ub,AyonBeato:1999rg,Bronnikov:2005gm,Berej:2006cc,Bronnikov:2012ch,Rinaldi:2012vy,Stuchlik:2014qja,Schee:2015nua,Johannsen:2015pca,Myrzakulov:2015kda,Fan:2016hvf,Sebastiani:2016ras,Toshmatov:2017zpr,Chinaglia:2017uqd,Frolov:2017dwy,Bertipagani:2020awe,Nashed:2021pah,Simpson:2021dyo,Franzin:2022iai,Chataignier:2022yic,Ghosh:2022gka,Khodadi:2022dyi,Farrah:2023opk,Fontana:2023zqz,Boshkayev:2023rhr,Luongo:2023jyz,Luongo:2023aib,Cadoni:2023lum,Giambo:2023zmy,Cadoni:2023lqe,Luongo:2023xaw,Sajadi:2023ybm,Javed:2024wbc,Ditta:2024jrv,Al-Badawi:2024lvc,Ovgun:2024zmt,Corona:2024gth,Bueno:2024dgm,Konoplya:2024hfg,Pedrotti:2024znu,Bronnikov:2024izh,Kurmanov:2024hpn,Bolokhov:2024sdy,Agrawal:2024wwt,Belfiglio:2024wel,Stashko:2024wuq,Faraoni:2024ghi,Konoplya:2024lch,Khodadi:2024efq,Calza:2024qxn} for an inevitably incomplete selection of examples, as well as Refs.~\cite{Ansoldi:2008jw,Nicolini:2008aj,Sebastiani:2022wbz,Torres:2022twv,Lan:2023cvz} for recent reviews on the subject.~\footnote{Another interesting possibility are gravastars, which are not RBHs in a strict sense~\cite{Mazur:2001fv,Mazur:2004fk,Mottola:2023jxl}.} While most of these metrics have been introduced on purely phenomenological grounds it is known that possible sources for several RBH metrics lie in theories of non-linear electrodynamics~\cite{Bronnikov:2017sgg,Ghosh:2021clx,Bronnikov:2021uta,Bronnikov:2022ofk,Bokulic:2023afx}.

As alluded to earlier, our interest in this work is to explore the possibility that primordial RBHs may play the role of DM. As a proof-of-principle in this sense we will establish constraints on $f_{\text{pbh}}$, the fraction of DM in the form of PRBHs, focusing on the asteroid mass window, whose extent we will show can be further extended. To the best of our knowledge, we are aware of seven works in this little explored direction~\cite{Easson:2002tg,Dymnikova:2015yma,Pacheco:2018mvs,Arbey:2021mbl,Arbey:2022mcd,Banerjee:2024sao,Davies:2024ysj}. Ref.~\cite{Dymnikova:2015yma} studied primordial BHs with de Sitter interiors as DM candidates, but considering the case where the DM is actually constituted of remnants from the evaporation process. Ref.~\cite{Pacheco:2018mvs} studied the thermodynamics of primordial regular BHs, focusing however on the case where they do not evaporate, and therefore did not study constraints on $f_{\text{pbh}}$. Ref.~\cite{Arbey:2021mbl} studied the evaporation of a loop quantum gravity-inspired BH, and weakened constraints on $f_{\text{pbh}}$ were reported in a later proceeding~\cite{Arbey:2022mcd} (which however does not appear to be widely known)., Ref.~\cite{Banerjee:2024sao} studied signatures of primordial BHs with magnetic charge, which could be (as is often but necessarily the case) regular. Finally, Refs.~\cite{Easson:2002tg,Davies:2024ysj} entertained the case where the evaporation times of primordial regular BHs are significantly longer than the standard Schwarzschild case, and did not therefore explicitly compute the evaporation constraints we instead study here. The aim of this pilot study and our companion paper~\cite{Calza:2024xdh} is instead to provide a more comprehensive investigation of primordial regular BHs, considering a more diverse set of metrics and investigating constraints on $f_{\text{pbh}}$ in detail.

In our work, we shall consider three different non-rotating RBH metrics, as discussed in more detail in the following subsections. The static, spherically symmetric space-times we investigate are a subset of the Petrov type-D class of metrics. In four-dimensional Boyer-Lindquist coordinates, their line elements take the following general form:
\begin{eqnarray}
ds^2 = -f(r)dt^2 + g(r)^{-1}dr^2 + h(r)d\Omega^2\,,
\label{eq:ds2}
\end{eqnarray}
where $d\Omega^2=d\theta^2 +\sin^2(\theta) d\phi^2$ is the metric on the 2-sphere. We also require our space-times to be asymptotically flat, which amounts to the following requirements:
\begin{eqnarray}
f(r) \xrightarrow{r \to \infty} 1\,, \quad g(r) \xrightarrow{r \to \infty} 1\,, \quad h(r) \xrightarrow{r \to \infty} r^2\,.
\label{eq:asflat}
\end{eqnarray}
In addition, as stated earlier, we require our space-times to be \textit{tr}-symmetric (the non-\textit{tr}-symmetric case is covered in a companion paper~\cite{Calza:2024xdh}), which imposes the following additional conditions:
\begin{eqnarray}
f(r) = g(r)\,, \quad h(r) = r^2\,,
\label{eq:tr}
\end{eqnarray}
implying that the coordinate $r$ is effectively the areal radius. With the conditions given by Eqs.~(\ref{eq:asflat},\ref{eq:tr}) imposed upon Eq.~(\ref{eq:ds2}), our most general line element therefore takes the following form:
\begin{eqnarray}
ds^2 = -f(r) dt^2 + f(r)^{-1}dr^2 + r^2 \left [ d\theta^2 +\sin^2(\theta) d\phi^2 \right ] \,.
\label{eq:ds2trsymmetric}
\end{eqnarray}
In what follows, we refer to the function $f(r)$ as being the ``metric function''. The three different RBH solutions we consider, which we will discuss very shortly in Sections~\ref{subsec:bardeen}--~\ref{subsec:cgsv}, are characterized by different functional forms of $f(r)$.

\begin{figure}
\centering
\includegraphics[width=1.0\columnwidth]{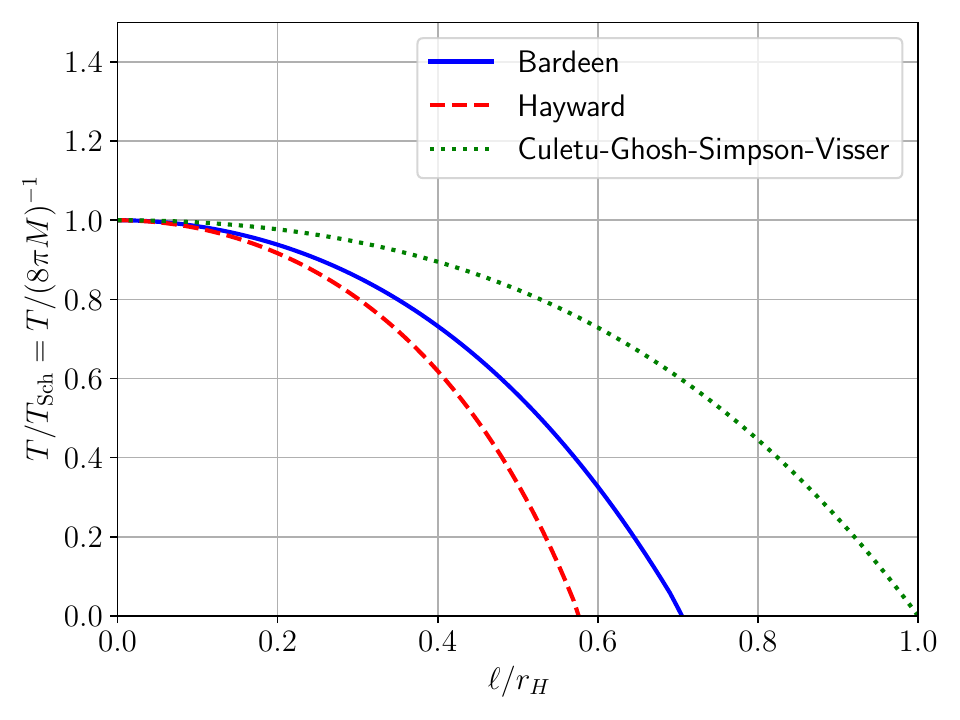}
\caption{Evolution of the temperatures (normalized to the temperature of Schwarzschild black holes, $T_{\text{Sch}}=1/8\pi M$) as a function of the regularizing parameter $\ell$ (normalized to the event horizon radius $r_H$) for the three regular black holes studied in this work: the Bardeen (blue solid curve, Sec.~\ref{subsec:bardeen}), Hayward (red dashed curve, Sec.~\ref{subsec:hayward}), and Culetu-Ghosh-Simpson-Visser (green dotted curve, Sec.~\ref{subsec:cgsv}) regular space-times. In all cases the temperature is a monotonically decreasing function of $\ell$. Note that the range of allowed values of $\ell/r_H$ is different for the three regular black holes.}
\label{fig:temperatures}
\end{figure}

An important parameter characterizing the behaviour of BHs is their temperature $T$. This is particularly crucial when evaluating evaporation constraints on PBHs, given that the temperature controls the strength of the emitted radiation, which in turn can be directly constrained by various observations. We treat the temperature of the RBHs as being the usual Gibbons-Hawking one, i.e.\ the one evaluated by Wick rotating the metric in the standard way and imposing regularity in the Euclidean period~\cite{Gibbons:1977mu}. The cyclic imaginary time $\to$ temperature identification is legitimate if one can formally identify the Euclidean action $e^{-S}$ with the Boltzmann factor $e^{-\beta H}$ in the partition function, as usually done in finite temperature quantum field theory: in turn, this can be done if one is assuming the standard Boltzmann-Gibbs distribution, but may not be the consistent if other entropies are assumed (see e.g.\ the recent discussion in Ref.~\cite{Lu:2024ppa}). Since, as we will reiterate later, the RBHs we study are introduced on phenomenological grounds and we remain agnostic as to their theoretical origin (which may in principle be rooted within alternative entropic frameworks), in what follows we assume the Boltzmann-Gibbs distribution, so that the temperature of the RBHs in question is the standard Gibbons-Hawking one, and is given by the following:
\begin{eqnarray}
T=\frac{\kappa}{2\pi}=\frac{f'(r)}{4\pi}\vert_{r_H}\,,
\label{eq:temperature}
\end{eqnarray}
where the prime indicates a derivative with respect to $r$, and $\kappa$ is the BH surface gravity, given by the following:
\begin{eqnarray}
\kappa=\frac{f'(r)}{2}\vert_{r_H}\,.
\label{eq:kappa}
\end{eqnarray}
In Eqs.~(\ref{eq:temperature},\ref{eq:kappa}), $r_H$ is the horizon radius, which is the solution to the following equation:
\begin{eqnarray}
g^{-1}(r_H)=f(r_H)=0\,,
\label{eq:rh}
\end{eqnarray}
with the first equality following from the choice of focusing on \textit{tr}-symmetric space-times. In the case of Schwarzschild BHs, where the metric function is $f(r)=1-2M/r$, one recovers the well-known expressions $r_H=2M$ and $T_{\text{Sch}}=1/8\pi M$. However, in more general space-times the horizon radius in Eq.~(\ref{eq:rh}) is not guaranteed to have a closed form expression, and the same therefore holds for the temperature in Eq.~(\ref{eq:temperature}). In Fig.~\ref{fig:temperatures} we show the evolution of the temperatures (normalized to the temperature of Schwarzschild BHs, $T_{\text{Sch}}=1/8\pi M$) of the three regular space-times we will introduce shortly, as a function of the regularizing parameter $\ell$ (itself normalized to the event horizon radius $r_H$). As the Figure clearly shows, for all three space-times the temperature is a monotonically decreasing function of $\ell$.~\footnote{This is actually not unrelated to the fact that most RBHs, and more generally hairy BH solutions, feature shadows whose size decreases relative to the size of the Schwarzschild BH shadow, i.e.\ $3\sqrt{3}M$, see Ref.~\cite{Vagnozzi:2022moj} for further more detailed discussions (see also Refs.~\cite{Kazakov:1993ha,EventHorizonTelescope:2021dqv,EventHorizonTelescope:2022xqj,Bronnikov:2023uuv,Vertogradov:2024dpa,Vertogradov:2024qpf,Vertogradov:2024jzj}). In fact, the equations governing BH temperature and shadow size bear some resemblance, so it is not unreasonable that an increase/decrease in one could lead to an increase/decrease in the other. However, general thermodynamical considerations lead to expect that well-behaved RBHs should decrease their temperature as they approach the extremal value of the regularizing parameter. We defer a study of RBHs which increase their temperature to a follow-up paper. In passing, we also note that Kerr BHs display an enhanced primary photon emission spectrum as the spin parameter is raised~\cite{Arbey:2019jmj,Arbey:2019vqx,Arbey:2020yzj}.} We also see that the temperatures of these BHs approach zero in the extremal limit. As has been observed earlier, in this limit the usual constraints on $f_{\text{pbh}}$ vanish, since these BHs do not evaporate~\cite{Pacheco:2018mvs}.

A final caveat is in order before discussing the RBH metrics we consider. The latter are all regular in the sense of having finite curvature invariants $R \equiv g^{\mu\nu}R_{\mu\nu}$, $R_{\mu\nu}R^{\mu\nu}$, and ${\cal K} \equiv R_{\mu\nu\rho\sigma}R^{\mu\nu\rho\sigma}$. However, a more stringent criterion for regularity is that of geodesic completeness, which does not necessarily imply finiteness of curvature invariants and viceversa. We recall that a space-time is geodesically complete if any geodesic thereon can be extended to arbitrary values of their affine parameter. In other words, all geodesics extend (or can be extended) for all times. A number of ``popular'' RBHs have indeed been shown to have finite curvature invariants but to be geodesically incomplete~\cite{Zhou:2022yio}. This includes the well-known Hayward RBH, which is among the ones we shall consider here. Nevertheless, given the significant interest in this metric, the fact that it is widely taken as prototype for RBHs, and our phenomenological goal of going beyond Schwarzschild PBHs, we will take this space-time into consideration, while cautioning the reader about the above issues, and therefore that the Hayward metric (but more generally all the metrics we will study) should be considered nothing more than phenomenological toy models at this stage. Note, in addition, that the stability of RBHs featuring inner horizons is currently a matter of debate in the literature~\cite{Carballo-Rubio:2021bpr,Carballo-Rubio:2022kad,Bonanno:2022jjp,Bonanno:2022rvo,Carballo-Rubio:2022twq}.

\subsection{Bardeen black hole}
\label{subsec:bardeen}

The Bardeen BH is easily one of the best known RBHs, and one of the first ones to ever have been proposed~\cite{Bardeen:1968ghw}. It is characterized by the following metric function:~\footnote{Note that in all the metrics considered, the parameter $M$ appearing in the metric function can always be unambiguously identified with the BH mass (either the Komar, ADM, Misner-Sharp-Hernandez, or Brown-York mass). This is important as the constraints we subsequently derive on $f_{\text{pbh}}$ as a function of $M_{\text{pbh}}$ will implicitly identify the latter with $M$.}
\begin{eqnarray}
f_{\text{B}}(r)=1-\frac{2Mr^2}{(r^2+\ell^2)^{3/2}}\,,
\label{eq:frbardeen}
\end{eqnarray}
where, in terms of the BH mass $M$, the regularizing parameter satisfies $\ell \leq \sqrt{16/27} M \sim 0.77\,M$ in order for the metric to describe a BH and not a horizonless object. Note that the Schwarzschild metric function is recovered in the $\ell \to 0$ limit. A perhaps physically more motivated choice is to express quantities in units of the horizon radius $r_H$, defined as the largest root of the equation $f(r_H)=0$, in which case the regularizing parameter is subject to the constraint $\ell \lesssim 0.70\,r_H$ (which is the point at which $T=0$ in Fig.~\ref{fig:temperatures}). In order to obtain this limit we have computed the solution to $f(r_H)=0$ fixing $M=1$, in order to extrapolate $r_H(\ell)$, before analyzing for which real values of the parameter $n$ the equation $\ell=nr_H(\ell)$ admits solutions. We note that the peculiar factor of $\sqrt{16/27}$ in the extremality condition appears when one demands that the cubic equation determining the horizon(s) location(s) of the Bardeen RBH, $f_B(r)=0$, admits one real non-zero root: with some algebraic manipulation, one sees that this condition amounts to the requirement that $\sqrt{27\ell^8M^4-16\ell^6M^6}$ is real, from which the requirement $\ell \leq \sqrt{16/27} M$ follows.

It is worth noting that the Bardeen BH possesses a de Sitter (dS) core which replaces the central singularity of the Schwarzschild BH. This is evident by noting that, in the limit $r \to 0$, the metric function goes as $f_B(r) \propto r^2$, exactly as expected for an asymptotically dS space-time. Although originally introduced on phenomenological grounds, it is now known that the Bardeen RBH can emerge from a magnetic monopole source~\cite{Ayon-Beato:2000mjt}, potentially within the context of a specific non-linear electrodynamics theory~\cite{Ayon-Beato:2004ywd}. Another possible origin for the Bardeen RBH are quantum corrections to the uncertainty principle~\cite{Maluf:2018ksj}. Irrespective of its origin, and consistently with the approach pursued for the other space-times, we consider this solution as a model-agnostic phenomenological toy model.

\subsection{Hayward black hole}
\label{subsec:hayward}

Another widely known regular space-time is the Hayward RBH~\cite{Hayward:2005gi}, characterized by the following metric function:
\begin{eqnarray}
f_{\text{H}}(r)=1-\frac{2Mr^2}{r^3+2M\ell^2}\,.
\label{eq:frhayward}
\end{eqnarray}
If expressed in terms of BH mass $M$, the regularizing parameter for the Hayward BH is subject to the same limit as that of the Bardeen BH, i.e.\ $\ell \leq \sqrt{16/27}\,,M$. On the other hand, if expressed in terms of the more physically motivated horizon radius, the limit is instead $\ell \lesssim 0.57\,r_H$ (again this is the point where $T=0$ in Fig.~\ref{fig:temperatures}). We note that the Schwarzschild metric function is recovered in the $\ell \to 0$ limit. The factor of $\sqrt{16/27}$ in the extremality condition originates from considerations similar to those we made for the Bardeen RBH in Sec.~\ref{subsec:bardeen}: with some algebraic manipulation, one sees that this condition amounts to the requirement that $\sqrt{27\ell^4M^2-16\ell^2M^4}$ is real, from which the requirement $\ell \leq \sqrt{16/27} M$ follows.

Just as the Bardeen RBH possesses a dS core, so does the Hayward RBH. Indeed, introducing a dS core characterized by a (positive) cosmological constant $\Lambda= 3/\ell^2$ in order to prevent the central singularity was precisely the original justification for the Hayward BH which, just like its Bardeen counterpart, was proposed on purely phenomenological grounds. Nevertheless, potential theoretical origins for the Hayward BH have been investigated, and range from corrections to the equation of state of matter at high density~\cite{Sakharov:1966aja,1966JETP...22..378G}, finite density and finite curvature proposals~\cite{1982JETPL..36..265M,1987JETPL..46..431M,Mukhanov:1991zn}, theories of non-linear electrodynamics~\cite{Kumar:2020bqf,Kruglov:2021yya}, and more generally as the result of corrections due to quantum gravity~\cite{Addazi:2021xuf,AlvesBatista:2023wqm}. Just as with the Bardeen RBH, we shall here treat the Hayward RBH as a model-agnostic phenomenological toy model for a singularity-free space-time.

\subsection{Culetu-Ghosh-Simpson-Visser black hole}
\label{subsec:cgsv}

The regular space-times considered so far featured dS cores, which in itself is a very common feature of several RBH metrics. Nevertheless, another interesting phenomenological possibility consists in considering ``hollow'' RBHs wherein the central singularity is replaced by an asymptotically Minkowski core, where the associated energy density and pressure asymptote to zero. This is quite unlike the case of the dS core where the energy density asymptotes to a finite value associated to a positive cosmological constant, and the pressure asymptotes to an equal but opposite value. Possible theoretical/mathematical motivations for considering RBHs with Minkowski cores include the fact that the vanishing energy density can significantly simplify the physics in the deep core, whereas the otherwise messy solutions to polynomial equations (which often cannot be written down in closed form) can be traded for arguably more elegant special functions, resulting in the space-time being more tractable. Our physical motivation in considering this class of BHs is instead to broaden the range of physical properties and phenomenological implications of PRBHs, going beyond the dS core RBHs studied thus far.

With this in mind, we consider a RBH featuring a Minkowski core, independently studied by Culetu~\cite{Culetu:2013fsa,Culetu:2014lca}, Ghosh~\cite{Ghosh:2014pba}, as well as Simpson and Visser~\cite{Simpson:2019mud}. Although such a RBH does not have any particular name associated to it in the literature, here we refer to it as CGSV BH (from the initials of the four authors above). The space-time is characterized by the following metric function:
\begin{eqnarray}
f_{\text{CGSV}}(r)= 1-\frac{2M}{r}\exp \left ( -\frac{\ell}{r} \right ) \,.
\label{eq:frcgsv}
\end{eqnarray}
The horizon radius $r_H$, for which a closed form expression is not available in the Bardeen and Hayward cases, here is given by:
\begin{eqnarray}
r_H=-\frac{\ell}{W \left ( -\frac{\ell}{2 M} \right ) }\,,
\label{eq:rhcgsv}
\end{eqnarray}
where $W$ denotes the Lambert function. Considering the principal branch $W_0$, a real and positive horizon radius is present for:
\begin{eqnarray}
W_0 \left ( -\frac{\ell}{2 M} \right ) \leq 0 \implies 0 \leq \ell < \frac{2M}{e}\,,
\label{eq:lambhert}
\end{eqnarray}
or, alternatively, $0 \leq \ell < r_H$. While the CGSV BH was original introduced purely on phenomenological/mathematical grounds, it was shown in Refs.~\cite{Kumar:2020ltt,Singh:2022xgi} that such a space-time can emerge within the context of GR coupled to a specific non-linear electrodynamics source. In this case, denoting by $g$ the non-linear electrodynamics coupling constant/charge, the regularizing parameter $\ell$ is given by $\ell=g^2/2M$, with $M$ the BH mass. Nevertheless, as with all the other RBHs considered, here we shall treat the CGSV RBH as a toy model for a regular space-time possessing a Minkowski core.

\section{Methodology}
\label{sec:methodology}

\subsection{Greybody factors}
\label{subsec:greybody}

A set of parameters playing a key role in describing the Hawking radiation spectra emitted from evaporating BHs are the so-called greybody factors (GBFs). These are functions of energy/frequency and angular momentum which govern the deviation of the emitted spectrum from that of a blackbody~\cite{Sakalli:2022xrb,Konoplya:2024lir,Konoplya:2024vuj}. Although the emitted Hawking radiation at the horizon takes the blackbody form, the potential barrier due to space-time geometry will attenuate the radiation, so that an observer at spatial infinity will measure a spectrum which differs from that of a blackbody by a frequency-dependent function $\Gamma(\omega)$. GBFs can be characterized by setting up a classical scattering problem around the BH potential barrier, with boundary conditions allowing for incoming wave packets from infinity or equivalently, due to the symmetries of the scattering problem, originating from the horizon. The scattering problem is governed by the so-called Teukolsky equation, which is a partial differential equation describing the propagation of perturbations of given spin in the BH background~\cite{Teukolsky:1973ha}.

For the static, spherically and \textit{tr}-symmetric metrics given by Eq.~(\ref{eq:ds2trsymmetric}) which we consider, the Teukolsky equation in spherical coordinates is separable. A key role in computing the GBFs is played by the radial Teukolsky equation, which we now report in full generality for the class of metrics in question. Using the Newman-Penrose (NP) formalism, the Teukolsky equation governing the evolution of (massless) perturbations of different spin can be condensed into a single master equation~\cite{Teukolsky:1973ha}:
\begin{align}
& \left [ - \frac{r^2}{f} \partial_t^2 + s \left ( r^2 \frac{f'}{f} -2 r \right ) \partial_t \right ] \Upsilon_s \nonumber \\
&+ \left [ (s+1) (r^2 f' + 2 r f) \partial_r \right ] \Upsilon_s \nonumber \\
&+\left [ \frac{1}{\sin{\theta}} \partial_\theta (\sin{\theta}\partial_\theta) + \frac{2 i s \cot{\theta}}{\sin{\theta}} \partial_\phi \right. \nonumber \\
&\left. + \frac{1}{\sin^2{\theta}} \partial^2_\phi -s -s^2 \cot^2{\theta} \right ] \Upsilon_s \nonumber \\
&+\left [ s r^2 f'' + 4 s r f' + 2 s f \right ] \Upsilon_s=0\,.
\label{eq:teukolsky}
\end{align}
Here, $\Upsilon_s$ represents a general perturbation of spin $s$, defined by the NP scalars relative to the respective perturbation. To not make the notation too heavy, we drop the $l$ and $m$ indices labelling the field mode, so $\Upsilon_s$ is understood to really mean $\Upsilon^{lm}_s$. We note that Eq.~(\ref{eq:teukolsky}) is separable if one makes the following wave ansatz:
\begin{eqnarray}
\Upsilon_s= \sum_{l,m} e^{-i \omega t } e^{i m \phi} S^{l}_s(\theta) R_s(r)\,,
\label{eq:upsilon}
\end{eqnarray}
where $\omega$ is the perturbation frequency, $l$ is the angular node number, and $m$ is the azimuthal node number.

The functions $S^l_s(\theta)$ contribute to defining the so-called spin-weighted spherical harmonics $S^s_{l,m}(\theta, \phi)=\sum S^l_s(\theta) e^{im\phi}$, satisfying the following equation~\cite{Fackerell:1977ghw,Suffern:1983ghw,Seidel:1988ue,Berti:2005gp}:
\begin{align}
&\left ( \frac{1}{\sin\theta}\partial_\theta(\sin\theta\,\partial_\theta)+\csc^2\theta\,\partial_\phi^2 \right. \nonumber \\
&\left. + \frac{2is\cot\theta}{\sin\theta}\partial_\phi+s-s^2\cot^2\theta+\lambda_l^s \right ) S_{l,m}^s=0\,,
\label{eq:slms}
\end{align}
where $\lambda_l^s\equiv l(l+1)-s(s+1)$ is the separation constant. For the spin 0 case, these functions reduces to the usual spherical harmonics $S_{l,m}^0=Y_{l,m}$.

Analogously to the Schwarzschild and Kerr BH cases~\cite{Teukolsky:1973ha}, the decoupled radial Teukolsky equation takes the following form~\cite{Harris:2003eg,Arbey:2021jif}:
\begin{align}
&\frac{1}{\Delta^s}\big(\Delta^{s+1}R'_s\big)' \nonumber \\
&+\left(\frac{\omega^2r^2}{f}+2i\omega sr-\frac{is\omega r^2f'}{f}+s(\Delta''-2)-\lambda_l^s\right)R_s=0\,,
\label{eq:radialteukolsky}
\end{align}
where $\Delta(r)\equiv r^2f(r)$ and $' \equiv \partial_r$. We set in purely ingoing boundary conditions, so the asymptotic solutions of Eq.~(\ref{eq:radialteukolsky}) are given by:
\begin{align}
& R_s \sim R^{\text{in}}_s \frac{e^{-i\omega r^{\star}}}{r}+ R^{\text{out}}_s \frac{e^{i\omega r^{\star}}}{r^{2s+1}} \quad (r\rightarrow \infty) \nonumber \\
& R_s \sim R^{\text{hor}}_s \Delta^{-s} e^{-i \omega r^{\star}} \quad \quad \quad \quad \, \, (r \rightarrow r_H)\,,
\label{eq:asymptotic}
\end{align}
where $r^{\star}$ is the tortoise coordinate defined by:
\begin{eqnarray}
\frac{dr^{\star}}{dr}=\frac{1}{f(r)}\,.
\label{eq:rstar}
\end{eqnarray}
We note that $r^{\star} \to r$ for large values of $r$, given that the metrics we consider are asymptotically flat.

In general, numerical integration methods are required to compute GBFs for general space-times, and this holds for our \textit{tr}-symmetric RBHs as well. In our work, we make use of the so-called shooting method (see Appendix~\ref{sec:appendix} for further details), which has already been successfully applied to these types of calculations in several earlier works~\cite{Rosa:2016bli,Rosa:2012uz,Calza:2021czr,Calza:2022ioe,Calza:2022ljw,Calza:2023rjt,Calza:2023gws,Calza:2023iqa}.

To begin with, we rewrite Eq.~(\ref{eq:radialteukolsky}) in terms of the rescaled coordinate $x$, given by the following:
\begin{eqnarray}
x \equiv \frac{r-r_H}{r_H}\,,
\label{eq:varx}
\end{eqnarray}
where $r_H$ is the largest real root of $f(r)=0$. With this substitution Eq.~(\ref{eq:radialteukolsky}) is rewritten as follows:
\begin{equation}
\begin{split}
& x^2(x+1)^3 f \ddot{R_s} \\ 
&+ (s+1) x(x+1) \left ( 2(x+1)f+(x+1)^2 \dot{f} \right ) \dot{R_s} \\
&+ V(\omega,x)R_s=0\,,
\end{split}
\label{eq:teukolsky2}
\end{equation}
where $\dot{ } \equiv \partial_x$, and $V(\omega,x)$ is given by:
\begin{equation}
\begin{split}
&V(\omega,x) = \left ( \frac{\omega^2 r_H^2 (x+1)^2}{f} + 2 i s (x+1) \omega \right. \\
& - i s r_H (x+1)^2 \frac{\dot f }{f} \omega + s \left ( 2 f + 4 (x+1) \dot f  \right.\\ 
&\left. + (x+1)^2 \ddot f  -2 \right ) -l(l+1) + s(s+1) \biggr) x (x+1)\,.
\end{split}
\end{equation}
In order to further simplify the problem, we work in units of horizon radius and therefore set $r_H=1$, so that $r=x+1$. In these units, the metric functions of the three RBHs under consideration are given by the following:
\begin{equation}
\begin{split}
& f_{\text{B}}(x)=1-\frac{(1+\ell^2)^{3/2}(x+1)^2}{ \left ( \ell^2+(x+1)^2 \right ) ^{\frac{3}{2}}}\,, \\
& f_{\text{H}}(x)=1-\frac{(x+1)^2}{(1-\ell^2) \left( (x+1)^3 - \frac{\ell^2}{\ell^2-1} \right ) } \,, \\
& f_{\text{CGSV}}(x)=1- \frac{e^{\ell - \frac{\ell}{x+1}}}{x+1}\,,\\
\end{split}
\end{equation}
for the Bardeen, Hayward, and CGSV space-times respectively.

Setting purely ingoing boundary in proximity of the horizon, the solutions to Eq.~(\ref{eq:teukolsky2}) can be expressed in the form of a Taylor expansion as follows~\cite{Wilson:1928ghw,Baberhasse:1935ghw,Leaver:1985ax,Leaver:1986vnb,Konoplya:2023ahd,Konoplya:2023ppx,Rosa:2012uz,Rosa:2016bli}:
\begin{eqnarray}
R_s(x)= x^{-s- \frac{i \omega}{\tau}} \sum_{n=0}^\infty a_n x^n\,,
\label{eq:near}
\end{eqnarray}
where $i \omega / \tau$ is a function of the field spin and the regularizing parameter, and also depends on the space-time in question. We refer the reader to Appendix~\ref{sec:appendix} for further details. 
The $a_n$ coefficients can be determined by substituting Eq.~(\ref{eq:near}) in Eq.~(\ref{eq:teukolsky2}) and iteratively solving the resulting algebraic equations. The near-horizon solution is then used to set the boundary conditions and numerically integrate the radial equation up to large distances, where the general form of the solution is the following:
\begin{eqnarray}
R(x) \xrightarrow{r \to \infty} R^{\text{in}}_s \frac{e^{-i \omega x}}{x}+R^{\text{out}}_s \frac{e^{i \omega x}}{x^{2s+1}}\,.
\label{eq:far}
\end{eqnarray}
The GBFs can then be computed from the $_s R^{l m}_{\text{in}}(\omega)$ coefficient. More specifically, the normalization of the scattering problem is set by requiring $a_0=1$. With this normalization, the GBFs then read:
\begin{eqnarray}
\Gamma^s_{l m}(\omega)=\delta_s \vert _s R^{l m}_{\text{in}}(\omega)\vert^{-2}\,,
\label{eq:gamma}
\end{eqnarray}
where the coefficient $\delta_s$ is given by:
\begin{eqnarray}
\delta_s = \tau \frac{ i e^{i \pi s} (2 \omega)^{2s-1} \Gamma \left ( 1-s- \frac{2 i \omega}{\tau} \right ) }{\Gamma \left ( s-\frac{2 i \omega}{\tau} \right ) } \,,
\label{eq:delta}
\end{eqnarray}
where $\Gamma$ is the $\Gamma$ function. Using the method discussed above, we compute the GBFs for perturbations of different spin on the backgrounds of the three regular metrics discussed earlier, for different values of the regularizing parameter $\ell$. In the specific case $s=1$, we have checked that calculating the GBFs up to $l=4$ is sufficient for our purposes. The GBFs we calculate are then used to characterize the Hawking evaporation spectra.

\begin{figure}
\centering
\includegraphics[width=1.0\columnwidth]{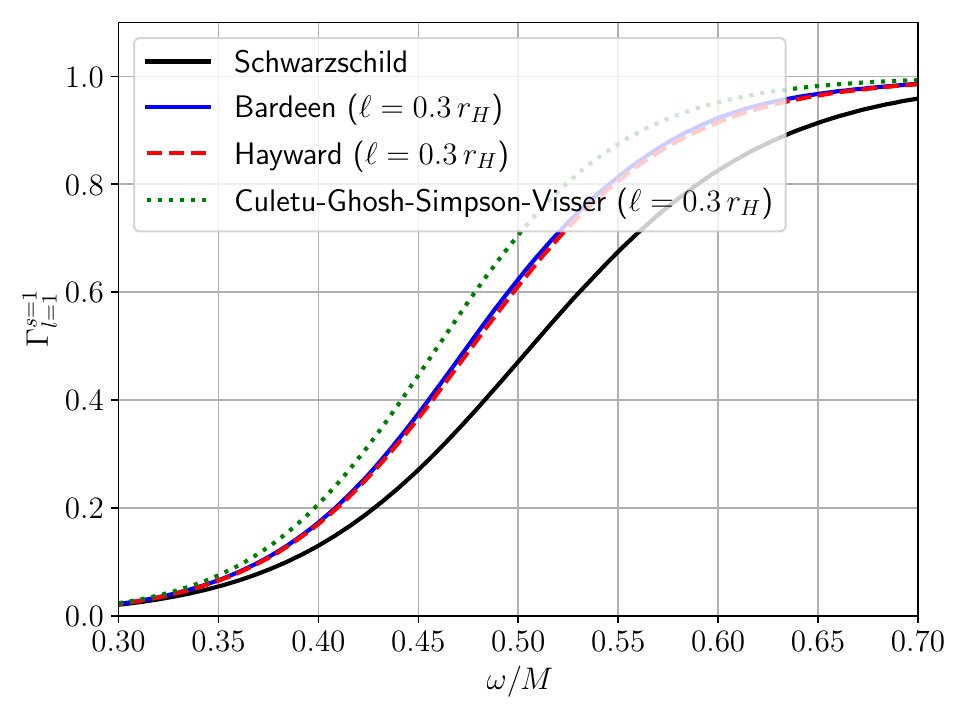}
\caption{Greybody factors $\Gamma_{l=1}^{s=1}$ as a function of $\omega/M$ for Schwarzschild BHs (black curve), as well as the Bardeen (blue solid curve), Hayward (red dashed curve), and Culetu-Ghosh-Simpson-Visser (green dotted curve) regular space-times. For illustrative purposes we only plot $\Gamma_{l=1}^{s=1}$, since we are interested in photons ($s=1$) and the dominant emission mode is the $l=1$ one. We have fixed the regularizing parameter to $\ell=0.3r_H$ for all three regular space-times. We see that in all three cases the GBFs are consistently (slightly) higher than their Schwarzschild counterparts. The features shown in this plot do not change sensibly for higher values of $l$ and other values of $\ell$.}
\label{fig:gbfs}
\end{figure}

In Fig.~\ref{fig:gbfs} we show the $\Gamma_{l=1}^{s=1}$ GBFs for the Schwarzschild BH and the three regular space-times we study. For illustrative purposes we have focused on the $\Gamma_{l=1}^{s=1}$ GBFs, since we are interested in photons ($s=1$) and the dominant emission mode is the $l=1$ one. We also note that, since we are considering spherically symmetric space-times, the $(2l+1)$ different $m$ modes are degenerate. We have fixed the regularizing parameter to $\ell=0.3r_H$ for all three regular space-times, also for illustrative purposes. We see that, for all three regular space-times, the GBFs are slightly higher than their Schwarzschild counterparts (by $\lesssim 20\%$ at most), and asymptote to the latter for both $\omega/M \lesssim 0.3$ and $\omega/M \gtrsim 0.7$. While for definiteness we have focused on these specific values of $s$, $l$, and $\ell$, we have explicitly checked that the features described above are present for other values of the parameters in question as well.

\subsection{Evaporation spectra}
\label{subsec:spectra}

We now discuss our computation of the photon spectra resulting from Hawking evaporation of the regular BHs discussed previously. In what follows, we only account for the primary photon spectrum. Nevertheless, we have checked that in the mass region of interest the impact of the secondary component of the spectrum, i.e.\ that resulting from the decay into photons of other unstable particles which are also produced during the evaporation process, is negligible.

The Hawking radiation rate (number of particles emitted per unit time per unit energy) of a given particle species $i$ with spin $s$, as a result of Hawking evaporation, is given by the following~\cite{Hawking:1975iha,Hawking:1975vcx,Page:1976df,Page:1976ki,Page:1977um}:~\footnote{This expression implicitly assumes that the particles emitted by the BH are not coupled to the regularizing parameter $\ell$, an assumption which is reasonable.}
\begin{eqnarray}\label{prim}
\frac{d^2N_i}{dtdE_i}=\frac{1}{2\pi}\sum_{l,m}\frac{n_i\Gamma^s_{l,m}(\omega)}{ e^{\omega/T}\pm 1}\,,
\label{eq:d2ndtdei}
\end{eqnarray}
where $n_i$ is the number of degrees of freedom of the particle in question, $\omega=E_i$ is the mode frequency (in natural units), $\Gamma^s_{l,m}$ are the GBFs discussed previously, and we have implicitly set $k_B=1$. Note that the plus (minus) sign in the denominator is associated to fermions (bosons). Following the methodology discussed in Sec.~\ref{subsec:greybody}, we calculate the GBFs within all the BH space-times in question for photons ($s=1$), up to $l=4$ (note that the angular node number $l$ should not be confused with the regularizing parameter $\ell$). We have checked that adding higher $l$ modes does not appreciably improve the resulting spectra.

\begin{figure}
\centering
\includegraphics[width=1.0\columnwidth]{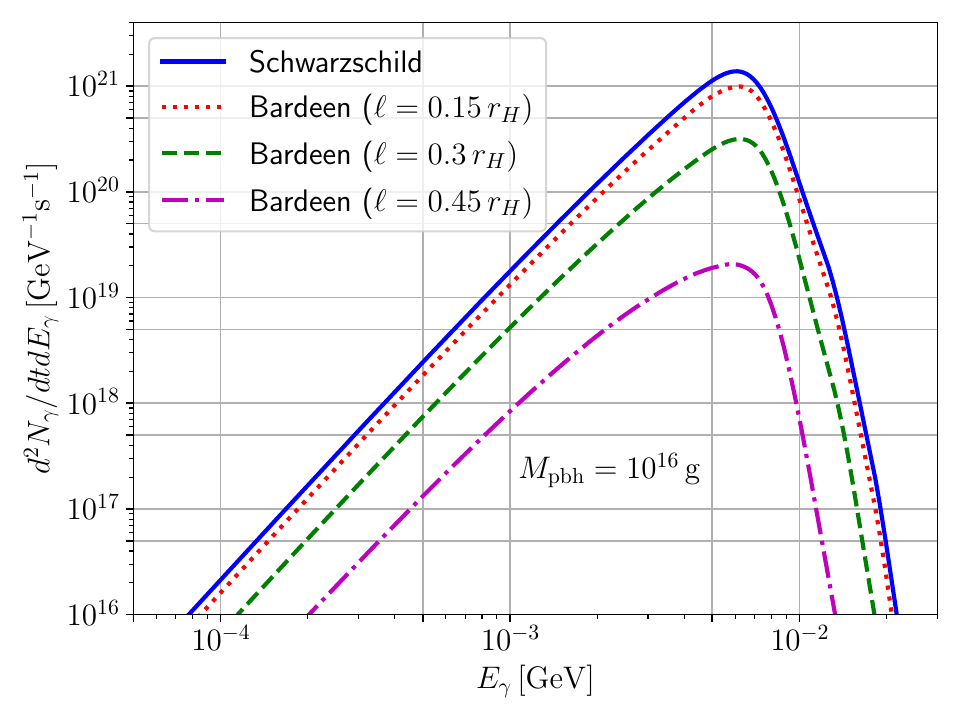}
\caption{Primary photon spectra resulting from the evaporation of Bardeen black holes of mass $10^{16}\,{\text{g}}$ for different values of the regularizing parameter $\ell$ (normalized by the horizon radius $r_H$): $\ell/r_H=0.15$ (red dotted curve), $0.3$ (green dashed curve), and $0.45$ (magenta dash-dotted curve). The blue solid curve corresponds to the case $\ell/r_H=0$, which recovers the Schwarzschild black hole.}
\label{fig:spectraphotonsbardeen}
\end{figure}

\begin{figure}
\centering
\includegraphics[width=1.0\columnwidth]{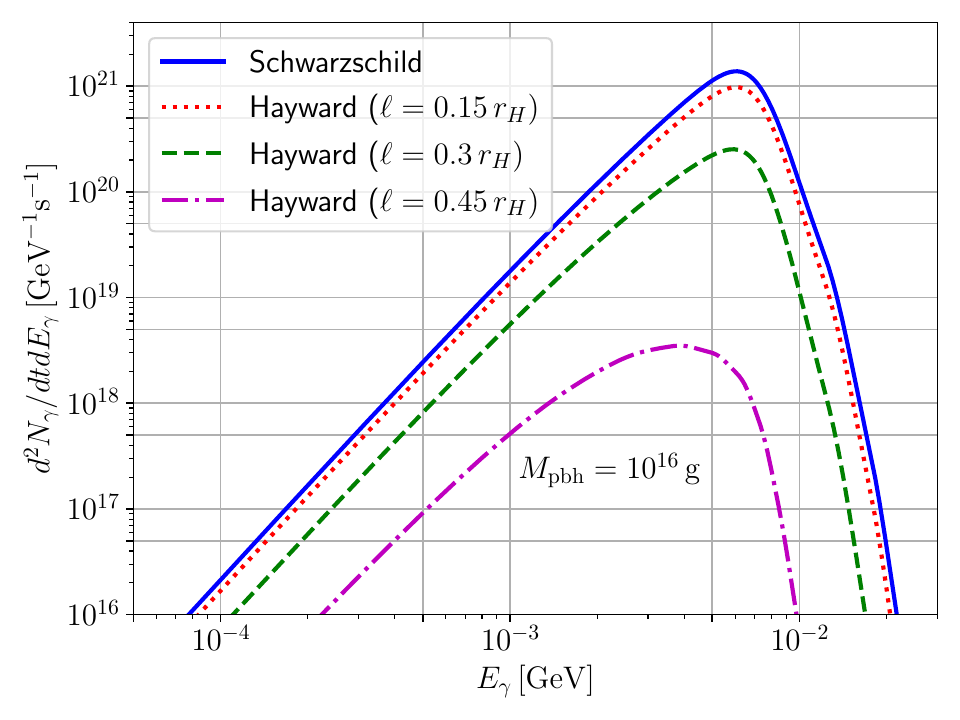}
\caption{As in Fig.~\ref{fig:spectraphotonsbardeen}, but for Hayward black holes, with identical values of the regularizing parameter $\ell/r_H$ and identical color coding.}
\label{fig:spectraphotonshayward}
\end{figure}

\begin{figure}
\centering
\includegraphics[width=1.0\columnwidth]{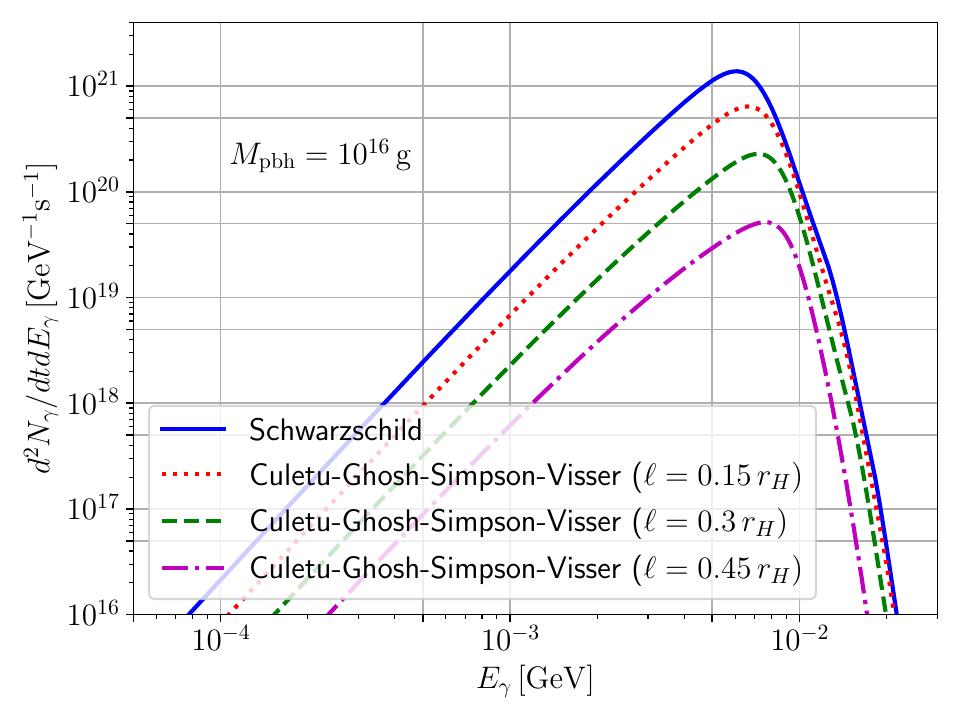}
\caption{As in Fig.~\ref{fig:spectraphotonsbardeen}, but for Culetu-Ghosh-Simpson-Visser black holes, with identical values of the regularizing parameter $\ell/r_H$ and identical color coding.}
\label{fig:spectraphotonscgsv}
\end{figure}

We show examples of the resulting evaporation spectra in Figs.~\ref{fig:spectraphotonsbardeen},~\ref{fig:spectraphotonshayward}, and~\ref{fig:spectraphotonscgsv}. The spectra obviously depend on the mass of the evaporating PRBH, which we have set to $M_{\text{pbh}}=10^{16}\,{\text{g}}$, as it sits roughly in the middle of the mass range of interest. Nevertheless, we stress that the features we discuss below do not depend on the chosen mass. The resulting spectra all peak approximately between $5\,{\text{MeV}}$ and $10\,{\text{MeV}}$.

For the Bardeen, Hayward, and CGSV PRBHs we observe that an increase in the regularizing parameter $\ell$ leads to a decrease in the intensity of the spectra \textit{at all energies}. This is na\"{i}vely what one could expect by inspecting the temperature evolution shown in Fig.~\ref{fig:temperatures}, since for all three these PRBHs the temperatures decrease relative to their Schwarzschild counterparts. However, as Eq.~(\ref{eq:d2ndtdei}) shows, the temperature is not the only quantity at play, since the GBFs enter as well. As we have seen in Fig.~\ref{fig:gbfs}, the GBFs for all three PRBHs are larger than their Schwarzschild counterparts, which would seem to counteract the effect of a lower temperature. A posteriori, however, the dominant effect turns out to be that of a lower temperature. We could actually have expected as much, since the resulting spectra given by Eq.~(\ref{eq:d2ndtdei}) are linear in the GBFs, but depend exponentially on the temperature. Therefore, one might expect that a decrease in the temperature will have a much more dramatic effect than a comparable increase in the GBFs. This is precisely what we observe in Figs.~\ref{fig:spectraphotonsbardeen},~\ref{fig:spectraphotonshayward}, and~\ref{fig:spectraphotonscgsv}.

Aside from the amplitude of the spectrum, for the Bardeen and Hayward PRBHs we note that the position of the peak in the spectrum is only mildly affected by the regularizing parameter, an increase in which pushes the peak towards slightly lower energies. On the other hand, for the CGSV BH an increase in the regularizing parameter pushes the peak towards slightly higher energies. We do not exclude that this different behaviour may be related to the type of core being considered, and we defer a more detailed investigation of this point to future work. At any rate, we expect that the behaviour of the spectra discussed above should lead to constraints on $f_{\text{pbh}}$ which are loosened for these classes of PRBHs.

\subsection{Evaporation constraints}
\label{subsec:constraints}

The spectra calculated in Sec.~\ref{subsec:spectra} are then used to set evaporation constraints on $f_{\text{pbh}}(M) \equiv \Omega_{\text{pbh}}/\Omega_{\text{dm}}$, the fraction of DM in the form of PBHs, where $\Omega_{\text{pbh}}$ and $\Omega_{\text{dm}}$ are the PBH and DM density parameters respectively. Specifically, the computed spectra are used to obtain predictions for the flux of photons resulting from Hawking evaporation, which are then directly compared against measurements of the extragalactic photon background across a wide range of energies (see e.g.\ Ref.~\cite{Auffinger:2022khh} for a recent review). Evaporation constraints are the dominant ones in the $10^{13}\,{\text{g}} \lesssim M_{\text{pbh}} \lesssim 10^{18}\,{\text{g}}$ mass range: the lower limit of the range is set by the requirement that PBHs have not yet evaporated at the time of recombination, whereas the upper limit is defined by measurements of the diffuse extragalactic $\gamma$-ray background (EGRB) in the energy range $100\,{\text{keV}} \lesssim E_{\gamma} \lesssim 5\,{\text{GeV}}$, given that the intensity of the Hawking radiation flux is inversely proportional to the mass of the evaporating BH. In what follows, we will direct our attention exclusively to PBHs for which $M_{\text{pbh}} \gtrsim 10^{15}\,{\text{g}}$: these have yet to fully evaporate today and, having formed deep during the radiation domination era, are therefore excellent non-baryonic DM candidates. There is another important reason for focusing on this mass range. As shown in Appendix~\ref{sec:appendixb}, PBHs (either Schwarzschild or the three PRBHs we consider) within this range have lifetimes much longer than the age of the Universe, are far from having fully evaporated at the present time, and have only lost a negligible fraction of their mass from formation until now. When using the symbol $M_{\text{pbh}}$, we are therefore safe in denoting the values of the PBH mass both at formation and now.

We work under the commonly adopted assumption that PBHs are isotropically distributed on sufficiently large scales. Therefore, the flux resulting from their evaporation and reaching us today is given by the \textit{redshifted} sum of the contributions from all evaporating PBHs in our Universe, and can be used to constrain the average extragalactic distribution of DM in the form of PBHs. Furthermore, we work within the (also commonly adopted) approximation of monochromatic mass distributions (which can be expected if the formation mechanism arises from an amplification of the power spectrum at a very specific scale), although the effect of extended mass distributions is the subject of active research~\cite{Kuhnel:2015vtw,Kuhnel:2017pwq,Carr:2017jsz,Raidal:2017mfl,Bellomo:2017zsr,Lehmann:2018ejc,Carr:2018poi,Gow:2019pok,DeLuca:2020ioi,Gow:2020cou,Ashoorioon:2020hln,Bagui:2021dqi,Mukhopadhyay:2022jqc,Papanikolaou:2022chm,Cai:2023ptf}. Finally, as discussed earlier, we only consider the primary photon contribution, as the secondary component resulting from the decay into photons of other unstable particles is verified a posteriori to be negligible given the mass range of interest. While all these are clearly approximations, albeit widely adopted ones, we are confident that they are appropriate given the aim of our work. Our main goal is in fact to examine how the limits on $f_{\text{pbh}}$ change when moving from the Schwarzschild PBH framework to that of the regular metrics presented in Sec.~\ref{sec:regular}, altering the asteroid mass window. It is more than reasonable to expect that the shift in constraints relative to the Schwarzschild case, $\delta f_{\text{pbh}}$, is only weakly affected by the above approximations. In other words we expect these approximations to have similar impacts on the constraints on $f_{\text{pbh}}$ relative to Schwarzschild BHs for the different RBHs discussed in Sec. \ref{sec:regular}, therefore leading to negligible effects on the relative shift $\delta f_{\text{pbh}}$, which is the quantity we are ultimately interested in. At any rate, the adopted approximations also allow for a more direct comparison to several previous works, and we therefore consider them appropriate for our pilot study, while stressing that their impact should definitely be explored in future follow-up works. Finally, note that we are tacitly assuming that PBHs cluster in the galactic halo in the same way as other forms of DM (unless they are extremely large, which is not the case for the mass range of interest).

In what follows, we therefore assume that PBHs all have the same initial mass $M_{\text{pbh}}$. Following Ref.~\cite{Carr:2009jm} we approximate the number of emitted photons in the logarithmic energy bin $\Delta E_{\gamma} \simeq E_{\gamma}$ as being given by $\dot{N}_{\gamma}(E_{\gamma}) \simeq E_{\gamma}(d\dot{N}_{\gamma}/dE_{\gamma})$. The emission rate of photons from Hawking evaporation per volume at a cosmological time $t$ is then given by~\cite{Carr:2009jm}:\footnote{In the original paper $M_{\text{pbh}}$ is assumed to be a function of time, due to the evaporation process. However, for the PBH masses considered in the present work ($M>10^{15}\,{\text{g}}$) we can safely assume $M$ to be roughly constant during the evaporation process~\cite{Carr:2021bzv}, see Appendix~\ref{sec:appendixb} for more detailed discussions.}
\begin{eqnarray}
\frac{dn_{\gamma}}{dt}(E_{\gamma},t) \simeq n_{\text{pbh}}(t)E_{\gamma}\frac{d^2{N}_{\gamma}}{dt dE_{\gamma}}(M_{\text{pbh}},E_{\gamma})\,,
\label{eq:dngammadt}
\end{eqnarray}
where $n_{\text{pbh}}(t)$ is the PBH number density at time $t$. By integrating and taking into account the redshift scaling of the photon energy and density we end up with:
\begin{align}
\begin{split}
&n_{\gamma 0}(E_{\gamma 0})\\
&= n_{\text{pbh}}(t_0) E_{\gamma 0}\int^{t_0}_{t_{\star}} dt(1 + z) \frac{d^2{N}_{\gamma}}{dt dE_{\gamma}}(M_{\text{pbh}},(1+z)E_{\gamma 0})\\
& = n_{\text{pbh}}(t_0) E_{\gamma 0}\int^{z_{\star}}_{0} \frac{dz}{H(z)}\frac{d^2{N}_{\gamma}}{dt dE_{\gamma}}(M_{\text{pbh}},(1+z)E_{\gamma 0})\,,
\label{eq:npbh1}
\end{split}
\end{align}
where $t_0$ denotes the present time, $t_{\star}$ and $z_{\star}$ are respectively the cosmic time and redshift at recombination, and $H(z)$ is the expansion rate. Finally, $n_{\gamma 0}(E_{\gamma 0})$ is the present number density of photons with energy $E_{\gamma 0}$. The resulting photon flux (more properly, the rate of photons per unit time per unit area per unit solid angle) is then given by:
\begin{eqnarray}
I(E_{\gamma 0}) \equiv \frac{c}{4\pi}n_{\gamma 0}(E_{\gamma 0})\,.
\label{eq:flux}
\end{eqnarray}
It is this quantity which can then be directly compared against observations.

We assume a spatially flat $\Lambda$CDM cosmological model in specifying the expansion rate entering into Eq.~(\ref{eq:npbh1}), with the same cosmological parameters as in Ref.~\cite{Carr:2009jm}. This allows us to robustly cross-check our Schwarzschild constraints on $f_{\text{pbh}}$ against those reported in the seminal Ref.~\cite{Carr:2009jm}, although we stress that our constraints are very stable against reasonable changes in the values of the cosmological parameters. Once the cosmological model is fixed, all the relevant quantities in Eq.~(\ref{eq:npbh1}) are known except for the present-day PBH number density, $n_{\text{pbh}}(t_0)$, which can be constrained from EGRB observations and is ultimately related to $f_{\text{pbh}}$. More specifically, for any given value of the PBH mass $M_{\text{pbh}}$, through Eqs.~(\ref{eq:d2ndtdei},\ref{eq:npbh1},\ref{eq:flux}) we can compute the unnormalized photon flux $I(E_{\gamma 0})/n_{\text{pbh}}(t_0)$, and adjust the normalization $n_{\text{pbh}}(t_0)$ by comparing against EGRB observations (as we will explain shortly). This procedures gives us an upper limit on $n_{\text{pbh}}(t_0)$, which can be translated into an upper limit on $f_{\text{pbh}}$ as follows:
\begin{eqnarray}
f_{\text{pbh}}(M_{\text{pbh}}) \equiv \frac{\Omega_{\text{pbh}}}{\Omega_{\text{dm}}} = \frac{n_{\text{pbh}}(t_0)M_{\text{pbh}}}{\rho_{\text{crit},0}\Omega_{\text{dm}}}\,,
\label{eq:npbhtofpbh}
\end{eqnarray}
where $\rho_{\text{crit},0}=3H_0^2/8\pi G$ is the present-day critical density, with $H_0$ the Hubble constant, and we recall that this procedure is done for various values of $M_{\text{pbh}}$.

We compare our theoretical predictions against various measurements of the EGRB. Specifically, we use observations of the EGRB from the HEAO-1 X-ray telescope in the $3$-$500\,{\text{keV}}$ range~\cite{Gruber:1999yr}, the COMPTEL imaging Compton $\gamma$-ray telescope in the $0.8$-$30\,{\text{MeV}}$ range~\cite{Schoenfelder:2000bu}, and the EGRET $\gamma$-ray telescope~\cite{Strong:2004ry}. A few comments are in order concerning the adopted datasets. While these are by now a couple of decades old, they basically still represent the state-of-the-art in the energy range of interest. One could entertain other observations, including local galactic measurements of the galactic $\gamma$-ray background~\cite{Carr:2016hva}, positron flux~\cite{Boudaud:2018hqb}, $0.511\,{\text{MeV}}$ annihilation radiation~\cite{DeRocco:2019fjq,Laha:2019ssq,Dasgupta:2019cae}, and various other sources. While these galactic observations could lead to potentially stronger limits, they depend strongly on the form of the PBH mass function (assumed to be monochromatic in our study), as well as the clustering properties of these PBHs. On the other hand, our limits on $f_{\text{pbh}}$ are effectively testing the average extragalactic distribution of DM. Finally, other measurements of the EGRB are available, e.g.\ from Fermi-LAT~\cite{Fermi-LAT:2014ryh}, but these are mostly important for energy ranges larger than the ones of interest, and therefore for PBHs lighter than the ones we are considering. Therefore, we believe the choice of datasets (which is the one adopted in several works estimating evaporation limits on PBHs) is appropriate given the objective of our study.

\begin{figure}
\centering
\includegraphics[width=1.0\columnwidth]{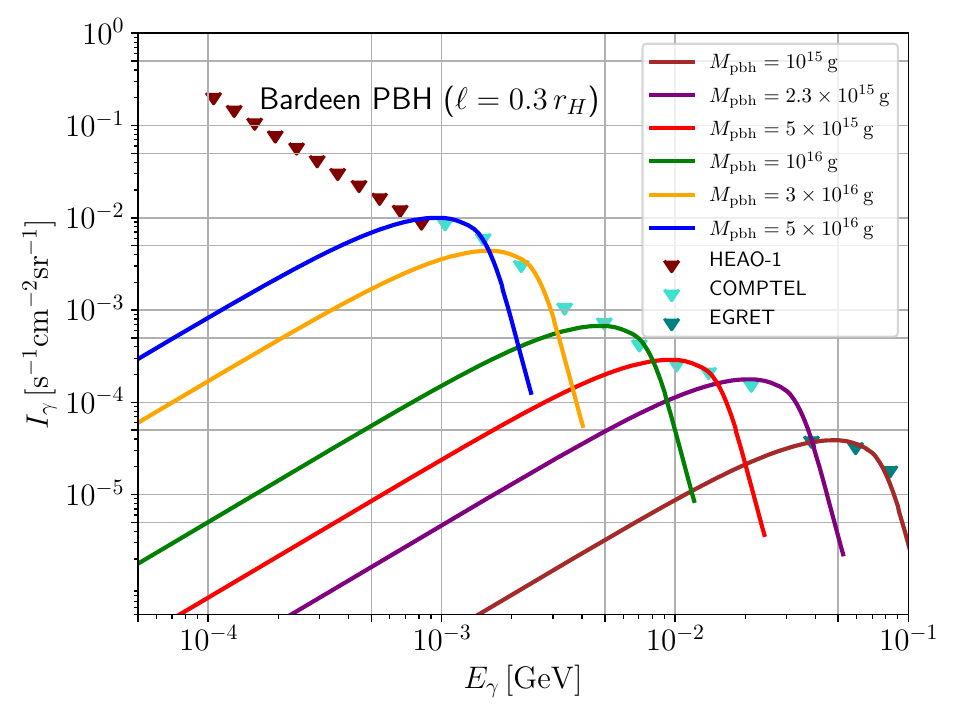}
\caption{Photon fluxes resulting from the evaporation of primordial Bardeen black holes with regularizing parameter $\ell=0.3r_H$, and masses of $M_{\text{pbh}}=10^{15}\,{\text{g}}$ (brown curve), $2.3 \times 10^{15}\,{\text{g}}$ (purple curve), $5 \times 10^{15}\,{\text{g}}$ (red curve), $10^{16}\,{\text{g}}$ (green curve), $3 \times 10^{16}\,{\text{g}}$ (green curve), and $5 \times 10^{16}\,{\text{g}}$ (blue curve). The triangles indicate $1\sigma$ upper limits on the extragalactic $\gamma$-ray background flux as measured by HEAO-1 (maroon), COMPTEL (turquoise), and EGRET (teal). For each value of $M_{\text{pbh}}$, the PBH fraction $f_{\text{pbh}}$ has been set to its upper limit, determined as soon as the first datapoint is overshot (the latter is different for different values of $M_{\text{pbh}}$).}
\label{fig:fluxdatalimits}
\end{figure}

To set upper limits on $n_{\text{pbh}}(t_0)$ -- and therefore $f_{\text{pbh}}$ through Eq.~(\ref{eq:npbhtofpbh}) -- we adopt the simple method first explained in the seminal Ref.~\cite{Carr:2009jm}, and later adopted in most of the works examining constraints on PBHs from the EGRB. Specifically, for each value of $M_{\text{pbh}}$, and for given values of the regularizing parameter $\ell$, the maximum allowed value of $f_{\text{pbh}}$ is determined by requiring that the predicted photon flux does not overshoot any of the ERGB datapoints by more than $1\sigma$. An example is shown in Fig.~\ref{fig:fluxdatalimits} for Bardeen PRBHs with regularizing parameter $\ell=0.3\,r_H$: for each of the mass values $M_{\text{pbh}}$ represented, the upper limit on $f_{\text{pbh}}$ is set as soon as the first datapoint is overshot. As is clear from the Figure, different PBH masses result in different datapoints being overshot. For each of the PRBHs considered, we use this procedure to determine upper limits on $f_{\text{pbh}}$ for fixed, representative values of $\ell$ ($\ell/r_H=0.15$, $0.3$, and $0.45$), comparing the results to the Schwarzschild case which is recovered when $\ell=0$.~\footnote{Clearly, from the statistical point of view, more precise analyses are possible. For instance, one could construct a metric to be minimized ($\chi^2$ or similar), or adopt a fully-fledged Bayesian approach exploring the joint $M$-$f_{\text{pbh}}$-$\ell$ posterior. Nevertheless, we believe our approach is sufficient for the purposes of our work, for several reasons. First and foremost, as with the approximations discussed earlier, adopting this method allows for a more direct comparison to several previous works. Furthermore, for most of these older datasets, often only the datapoints shown in Fig.~\ref{fig:fluxdatalimits} are available, with no further available details on aspects which would be required to properly build a $\chi^2$ or likelihood (e.g.\ correlation between the datapoints, instrumental details, and so on). Finally, we expect that the relative shift in $f_{\text{pbh}}$ limits with respect to their Schwarzschild counterparts will be largely unaffected by the adopted methodology. For all these reasons, and especially being ours a pilot study, we believe the adopted methodology is appropriate for the purposes of our work.} We note that the exact origin of the EGRB is currently a matter of debate~\cite{Dwek:2012nb}: although it is commonly believed that distant astrophysical sources such as blazars give a major contribution to the EGRB, there is no complete consensus on the level of this contribution. In this light, our approach of simply requiring that the PRBH evaporation contribution to the EGRB does not exceed any observed datapoint is rather conservative (given that there could in principle be a PBH contribution to the EGRB, should it be conclusively determined that known astrophysical sources are unable to fully account for the latter).

\section{Results}
\label{sec:results}

\begin{figure}
\centering
\includegraphics[width=1.0\columnwidth]{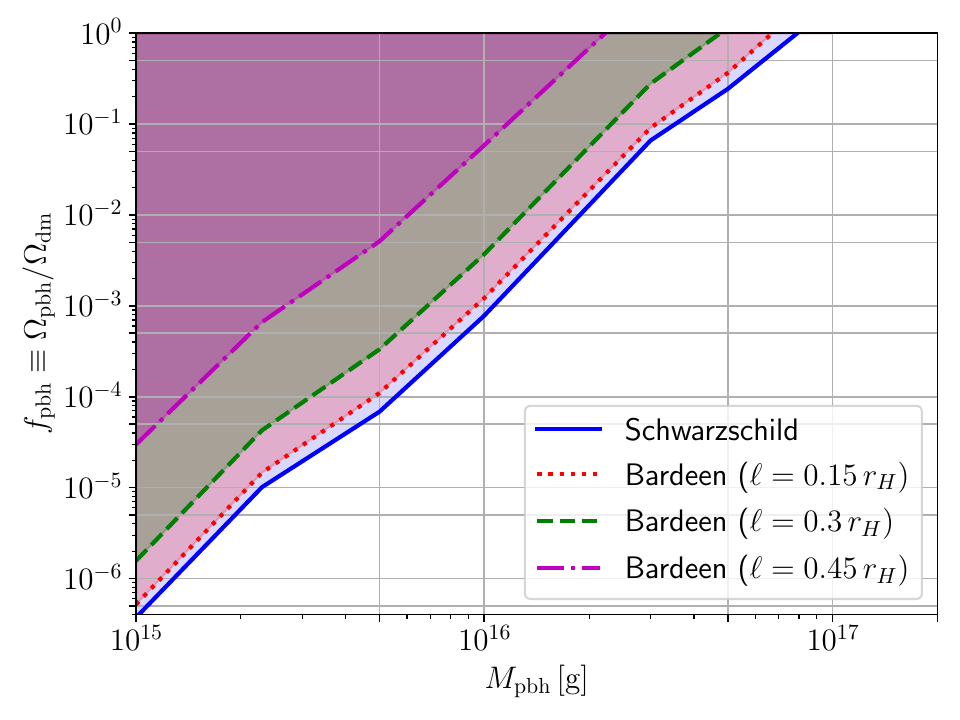}
\caption{Upper limits on $f_{\text{pbh}}$, the fraction of dark matter in the form of primordial regular Bardeen black holes, as a function of the black hole mass $M_{\text{pbh}}$. The limits are derived for different values of the regularizing parameter $\ell$ (normalized by the horizon radius $r_H$), with the shaded regions excluded: $\ell/r_H=0.15$ (red dotted curve), $0.3$ (green dashed curve), and $0.45$ (magenta dash-dotted curve). Note that the blue solid curve corresponds to the case $\ell/r_H=0$, which recovers the Schwarzschild black hole, whereas the value of $M_{\text{pbh}}$ corresponding to the upper right edge of the $f_{\text{pbh}}$ constraints marks the lower edge of the asteroid mass window.}
\label{fig:fpbhlimitsbardeen}
\end{figure}

\begin{figure}
\centering
\includegraphics[width=1.0\columnwidth]{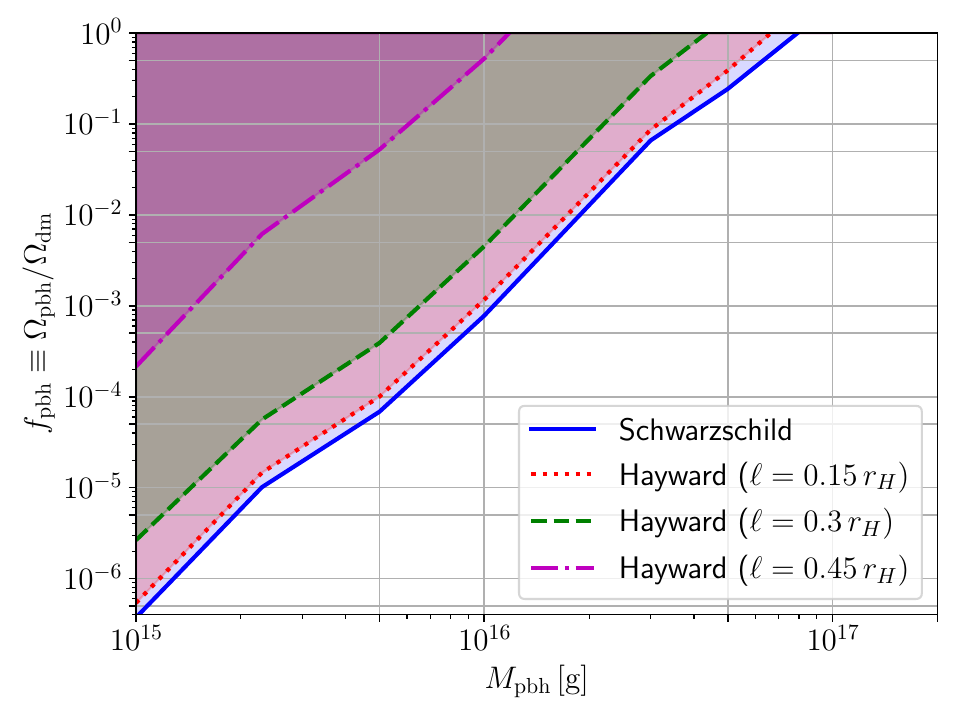}
\caption{As in Fig.~\ref{fig:fpbhlimitsbardeen}, but for primordial regular Hayward black holes, with identical values of the regularizing parameter $\ell/r_H$ and identical color coding.}
\label{fig:fpbhlimitshayward}
\end{figure}

\begin{figure}
\centering
\includegraphics[width=1.0\columnwidth]{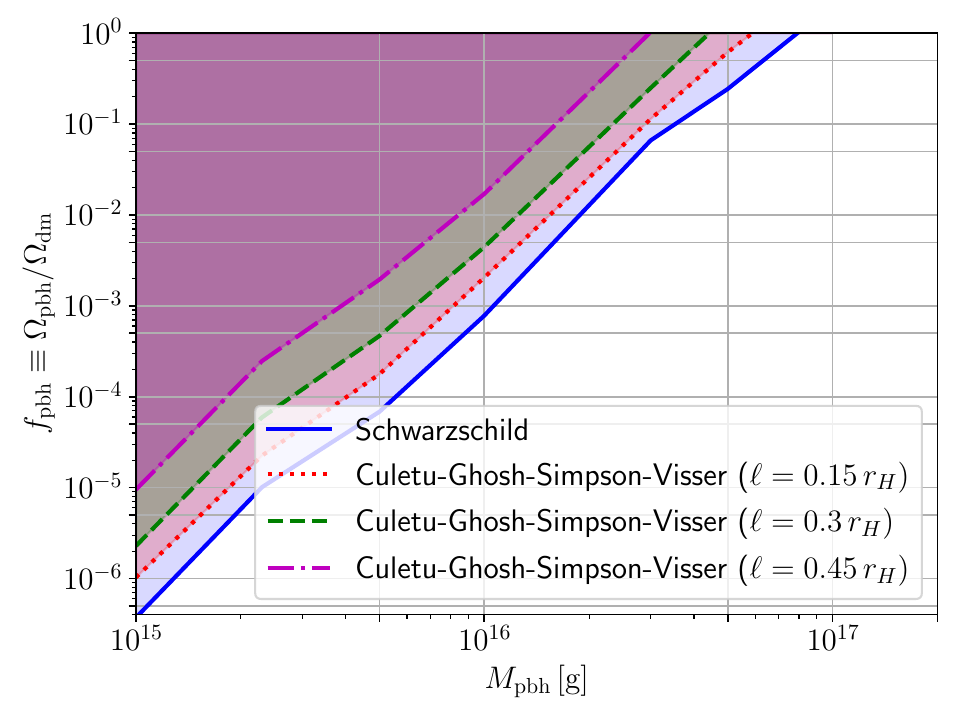}
\caption{As in Fig.~\ref{fig:fpbhlimitsbardeen}, but for primordial regular Culetu-Ghosh-Simpson-Visser black holes, with identical values of the regularizing parameter $\ell/r_H$ and identical color coding.}
\label{fig:fpbhlimitscgsv}
\end{figure}

For each of the PRBHs discussed in Sec.~\ref{sec:regular}, we now proceed to derive upper limits on $f_{\text{pbh}}$ as a function of the PRBH mass $M_{\text{pbh}}$, for different values of $\ell$, using the methodology presented in Sec.~\ref{subsec:constraints}. The results are shown in Figs.~\ref{fig:fpbhlimitsbardeen}, ~\ref{fig:fpbhlimitshayward}, and~\ref{fig:fpbhlimitscgsv} for Bardeen, Hayward, and CGSV PRBHs respectively. For each case, we also plot the constraints on $f_{\text{pbh}}$ for the $\ell=0$ case (blue solid curve in all the Figures), which correspond to the standard Schwarzschild PBH scenario widely studied in the literature. As a sanity check, we have verified that our $\ell=0$ constraints exactly recover those of the seminal Ref.~\cite{Carr:2009jm}. It is worth noting that, for any given value of $\ell$, the value of $M_{\text{pbh}}$ corresponding to the upper right edge of the $f_{\text{pbh}}$ constraints (i.e.\ the value of $M_{\text{pbh}}$ for which the limit reads $f_{\text{pbh}}<1$) marks the lower edge of the asteroid mass window.

For all PRBHs, we saw earlier that the temperature and photon spectra decrease in intensity with increasing regularizing parameter $\ell$ (see the discussion in Sec.~\ref{subsec:spectra}, and Figs.~\ref{fig:spectraphotonsbardeen}--\ref{fig:spectraphotonscgsv}). As we could have expected, this behaviour leads to overall looser constraints on $f_{\text{pbh}}$ (for any given $M_{\text{pbh}}$) relative to the standard limits reported for Schwarzschild PBHs in the literature. In the case of near-extremal Hayward PRBHs ($\ell=0.45r_H$) this behaviour is somewhat enhanced compared to the near-extremal Bardeen and CGSV PRBHs, with the upper limits on $f_{\text{pbh}}$ approximately three orders of magnitude looser than the corresponding Schwarzschild ones: again, this is somewhat unsurprising when comparing Fig.~\ref{fig:spectraphotonshayward} to Figs.~\ref{fig:spectraphotonsbardeen} and~\ref{fig:spectraphotonscgsv}. This could also have been expected from Fig.~\ref{fig:temperatures}, noting that the temperature of Hayward BHs decreases more rapidly with increasing $\ell/r_H$ relative to the Bardeen and especially CGSV ones. Although the temperature is not the only factor at play in determining the resulting evaporation spectra, given that the GBFs also play a key role as per Eq.~(\ref{eq:d2ndtdei}), it is reassuring to see that the temperature behaviour observed in Fig.~\ref{fig:temperatures} is qualitatively reflected in the limits on $f_{\text{pbh}}$ we derive.

As a result of the shifts discussed above, the lower edge of the asteroid mass window where PBHs could make up the entire DM component is modified for all three metrics considered. We recall that in the Schwarzschild case, the lower edge of the window lies at $M_{\text{pbh}} \simeq 10^{17}\,{\text{g}}$. For the PRBHs we consider, the looser constraints on $f_{\text{pbh}}$ result in the asteroid mass window further opening up by up to half a decade in mass or more. The maximum extension of the window is reached for the Hayward PRBH closer to extremality, in which case the lower edge decreases by about an order of magnitude to $M_{\text{pbh}} \simeq 10^{16}\,{\text{g}}$. Overall, we therefore observe that considering PRBHs in place of the standard Schwarzschild ones can relax the resulting constraints on $f_{\text{pbh}}$, further opening up the asteroid mass window. The allowed region for the window lower edge spans over a decade in mass, at least for the PRBHs and range of $\ell$ considered. We note that, as $\ell$ moves towards the extremal limit, the temperatures of these PRBHs $T$ approaches zero. This is indeed generally expected from thermodynamical arguments. In this case, which is the one studied in Ref.~\cite{Pacheco:2018mvs}, the PRBHs do not evaporate, and therefore our evaporation constraints do not apply. See also Refs.~\cite{Davies:2024ysj} for a related study appearing after ours.

Three comments are in order before concluding. Firstly we note that, for a given PRBH space-time, the curves describing the $f_{\text{pbh}}(M)$ limits are approximately, but not exactly parallel to the Schwarzschild ones (blue solid curves in Figs.~\ref{fig:fpbhlimitsbardeen}--\ref{fig:fpbhlimitscgsv}). The reason is simply that, as $\ell$ is increased, the datapoint shown in Fig.~\ref{fig:fluxdatalimits} which is first being overshot and therefore responsible for determining the $f_{\text{pbh}}$ limit can potentially change (in part due to the spectrum slightly changing shape).

Next, the constraints we have determined on $f_{\text{pbh}}$ at a fixed value of $\ell/r_H$ implicitly assume that all PRBHs in the Universe carry the same value of ``hair'' parameter $\ell$. However, particularly given our agnostic stance with regards to the origin of these space-times, in principle the value of $\ell/r_H$ can vary from PRBH to PRBH. To make an analogy, let us assume for a moment that Reissner-Nordstr\"{o}m BHs are astrophysically relevant. Then, since the electric charge $Q$ is not tied to a universal parameter of the underlying Einstein-Maxwell Lagrangian, there is no reason to expect it to carry the same value across all BHs. In the language of Ref.~\cite{Vagnozzi:2022moj}, the regularizing parameter for all three PRBHs considered is a ``specific hair'' rather than an ``universal hair'' (see Ref.~\cite{Vagnozzi:2022moj} for various examples of BH solutions carrying universal hair), unless one were able to tie $\ell$ to some fundamental parameter of the underlying theory, which however is not the case in the phenomenological approach we are following. In principle one should therefore account for the (non-monochromatic) $\ell$ distribution for PRBHs across the Universe to determine constraints on $f_{\text{pbh}}$. We see no obvious way of doing this, while noting that such a procedure would most likely result in upper limits on $f_{\text{pbh}}$ lying between the Schwarzschild and extremal cases: this observation suffices for our pilot study, and we defer a more complete investigation to future work.

Our final comment concerns the fact that evaporation limits on the PBH abundance are not the only ones at play. Indeed, as recently summarized in Ref.~\cite{Carr:2021bzv}, there are essentially four classes of limits, each of which is relevant in a different mass range: evaporation, lensing, dynamical, and accretion constraints. Constraints from the accretion of background gas at early times are relevant in a completely different mass range ($10^{33} \lesssim M_{\text{pbh}}/{\text{g}} \lesssim 10^{40}$ -- see Fig.~7 of Ref.~\cite{Carr:2021bzv} and Fig.~10 of Ref.~\cite{Carr:2020gox}). Although these have been derived assuming Schwarzschild PBHs, moving to the PRBH picture we have considered will not shift the relevant mass range by the $\gtrsim 18$ orders of magnitude required for these constraints to compete with the evaporation ones, unless the physics of gas accreting around RBHs changes drastically with respect to the standard picture, which appears very unlikely. Dynamical constraints, most of which are associated to the destruction of different astronomical objects by the passage of nearby PBHs, are also relevant in a completely different mass range ($10^{34} \lesssim M_{\text{pbh}}/{\text{g}} \lesssim 10^{55}$ -- see Fig.~7 of Ref.~\cite{Carr:2021bzv} and Fig.~10 of Ref.~\cite{Carr:2020gox}), and considerations completely analogous to those we made for accretion constraints hold.~\footnote{Potential exceptions to this mass range for dynamical constraints are those from capture of PBHs by white dwarfs or neutron stars at the centres of globular clusters~\cite{Capela:2012jz,Capela:2013yf,Pani:2014rca}, or from supernovae explosions resulting from transit of a PBH through a white dwarf~\cite{Graham:2015apa}. However, these limits are highly disputed because of uncertainties in the dark matter density in globular clusters~\cite{Ibata:2012eq,Defillon:2014wla}, or based on the results of hydrodynamical simulations~\cite{Montero-Camacho:2019jte}. For these reason, we will not consider the previously mentioned limits in our discussion.}

Of potentially more relevance to the present work are lensing constraints, which constrain the abundance of PBHs (and more generally massive compact halo objects) with masses $M_{\text{pbh}} \gtrsim 10^{15}\,{\text{g}}$. Indeed, it is lensing constraints which locate the upper edge of the asteroid mass window where PBHs can make up all the DM. Nevertheless, we expect that these constraints should not change when moving from the Schwarzschild PBH framework to the PRBHs considered in this work. Indeed, with all other quantities being fixed (mass of source, relative distances, and so on), lensing constraints only depend on the lens mass $M$, and are unaffected by the metric structure of the lens. Therefore, at fixed mass $M$, we can assume that the lensing limits on Schwarzschild PBHs hold for our PRBHs as well. Note that, as already pointed out in footnote~2, the parameter $M$ appearing in the RBH metrics can be unambiguously identified with the RBH mass, just as with the parameter $M$ in the Schwarzschild metric. We can therefore conclude that for the PRBHs we are considering it should only be the lower edge of the asteroid mass window which is altered with respect to the Schwarzschild case, but not the upper edge. In other words, space-times for which the lower edge moves towards lower masses (as in the Bardeen, Hayward, and CGSV PRBH cases) genuinely correspond to an enlarged asteroid mass window. Therefore, the window where Bardeen, Hayward, and CGSV PRBHs could account for all the DM is much larger compared to that of Schwarzschild PBHs.

Other potentially relevant constraints come from $\mu$-distortions in the Cosmic Microwave Background, and gravitational waves (either a stochastic background due to a population of coalescing PBHs or produced via second-order tensor perturbations generated by the scalar perturbations producing the PBHs, or associated to resolved events). The latter are expected to be relevant in a much higher mass range (again, see Fig.~7 of Ref.~\cite{Carr:2021bzv} and Fig.~10 of Ref.~\cite{Carr:2020gox}), whereas the former are somewhat dependent on the PBH formation scenario from high-$\sigma$ tails of density fluctuations, and in particular on the shape of the tail. At any rate, while the focus in the present pilot study has been solely on evaporation constraints from the ERGB, revisiting all these other important sources of constraints (including the ones we discussed earlier) is a worthwhile endeavour which we plan to explore in upcoming works.

\section{Conclusions}
\label{sec:conclusions}

Over the past decade, primordial black holes have regained tremendous interest as viable dark matter candidates, with the so-called ``asteroid mass window'' ($10^{17}\,{\text{g}} \lesssim M_{\text{pbh}} \lesssim 10^{23}\,{\text{g}}$) where PBHs could potentially account for the entire DM currently still open. Nearly all works on PBHs assume that these are Schwarzschild or Kerr BHs. However, while phenomenologically perfectly valid, such an assumption may stir some unease on the theoretical side, due to the appearance of singularities in these metrics. In our work, we have conducted a pilot study aimed at addressing a question which naturally merges the DM and singularity problems, arguably two among the most important open problems in theoretical physics: ``\textit{What if PBHs are non-singular}''? Our study of primordial regular BHs (PRBHs) has focused on three so-called \textit{tr}-symmetric metrics (including the well-known Bardeen and Hayward space-times), whereas our companion paper~\cite{Calza:2024xdh} considers non-\textit{tr}-symmetric metrics, including various metrics inspired from loop quantum gravity.

We show that evaporation constraints on $f_{\text{pbh}}$, the fraction of DM in the form of PRBHs, can be substantially loosened when moving away from the Schwarzschild picture, leading to the asteroid mass window further opening up. For the three PRBHs considered (the Bardeen, Hayward, and CGSV ones) the lower edge of the asteroid mass window is shifted by a decade in mass or more, leading to a larger region of parameter space where PRBHs could account for the entire DM component, which should be the target of the same probes proposed to test the standard window~\cite{Ray:2021mxu,Kainulainen:2021rbg,Ghosh:2022okj,Branco:2023frw,Amaral:2023ekd,Bertrand:2023zkl,Tran:2023jci,Dent:2024yje,Tamta:2024pow}. The nature of the regular BH core (de Sitter or Minkowski) does not appear to play a significant role in this sense. Overall, we have shown that the phenomenology of primordial regular BHs can be particularly rich, making the associated simultaneous solution to the DM and singularity problems one worthy of further study. We note that the constraints we obtained would become even weaker had we moved closer to the extremal limit for the regularizing parameter $\ell$: indeed, in this limit the temperatures of the RBHs we considered approaches zero, implying that there are no constraints from evaporation~\cite{Pacheco:2018mvs}.

We remark that the present work (alongside our companion paper~\cite{Calza:2024xdh}) should be intended as a pilot study, and there are a huge number of interesting follow-up directions. One interesting avenue for further work involves systematically revisiting other sources of constraints which have been studied in the Schwarzschild PBH case, including but not limited to lensing, accretion, and dynamical constraints: while we have argued that these should not alter our considerations on the asteroid mass window, a detailed study which would allow us to extend our constraints over a much larger region of $M_{\text{pbh}}$-$f_{\text{pbh}}$ plane is nevertheless in order. In addition, the metrics we have considered are inherently phenomenological in nature, and it would therefore be worth extending our study to non-singular metrics which enjoy a strong theoretical motivation (our companion paper~\cite{Calza:2024xdh} goes partially in this direction), including potentially metrics which are coupled to the cosmological expansion~\cite{Farrah:2023opk,Cadoni:2023lum,Cadoni:2023lqe,Faraoni:2024ghi,Calza:2024qxn}. Moreover, the formation mechanisms for these PRBHs is likely to be much more complex than the corresponding mechanisms for Schwarzschild PBHs: in fact, the space-times considered are not vacuum solutions of GR, which implies that Birkhoff's theorem does not (necessarily) hold, and correspondingly the endpoint of gravitational collapse may not be unique. At this stage it is impossible for us to make definitive statements on the matter without abandoning our agnostic viewpoint and assuming a specific underlying theoretical model, but the issue of PRBH formation (possibly within the context of inflationary models leading to an enhanced spectrum of curvature fluctuations over specific scales) is one which requires further study. Last but definitely not least, if PBHs are truly regular, one would hope to ascertain this via signatures complementary to those we have studied: gravitational wave observations, VLBI imaging, motion around BHs, or energy extraction are potentially interesting observables in this sense. For instance, the shadows of Bardeen, Hayward, and CGSV BHs are smaller than their Schwarzschild counterparts by up to $20\%$~\cite{Vagnozzi:2022moj}, whereas the motion of stars around these BHs (take for instance the S2 star orbiting around Sgr A$^{\star}$) would be altered compared to the motion around a Schwarzschild BH. These probes are being actively studied within the community as a means of distinguishing Schwarzschild BHs from alternatives thereto, and are likely to provide a promising route towards testing the regular nature of PRBHs, in the event that these are detected in the future and demonstrated to make up the DM, or even just a fraction thereof. We plan to address these and other related points in follow-up work.

\begin{acknowledgments}
\noindent We acknowledge support from the Istituto Nazionale di Fisica Nucleare (INFN) through the Commissione Scientifica Nazionale 4 (CSN4) Iniziativa Specifica ``Quantum Fields in Gravity, Cosmology and Black Holes'' (FLAG). M.C. and S.V. acknowledge support from the University of Trento and the Provincia Autonoma di Trento (PAT, Autonomous Province of Trento) through the UniTrento Internal Call for Research 2023 grant ``Searching for Dark Energy off the beaten track'' (DARKTRACK, grant agreement no.\ E63C22000500003). This publication is based upon work from the COST Action CA21136 ``Addressing observational tensions in cosmology with systematics and fundamental physics'' (CosmoVerse), supported by COST (European Cooperation in Science and Technology).
\end{acknowledgments}

\appendix

\section{Details on the computation of greybody factors}
\label{sec:appendix}

Here we provide a few more details on the computation of GBFs. We recall that we expressed the solutions to the radial Teukolsky equation, Eq.~(\ref{eq:teukolsky2}), in the form of a Taylor expansion as given by Eq.~(\ref{eq:near}). This is also known as a Frobenius series, being a by-product of a method for solving second-order differential equations named after Frobenius. The method applies to equations which take the following form
\begin{eqnarray}
u'' +p(x) u' + q(x) u=0\,,
\label{eq:frobenius}
\end{eqnarray}
in proximity of its singular points, namely those where $p(x)$ and/or $q(x)$ diverge. One can notice that Eq.~(\ref{eq:teukolsky2}) can be rewritten in the form of Eq.~(\ref{eq:frobenius}), with one of its singular point being at $x=0$, i.e.\ at the event horizon.

To solve the radial Teukolsky equation we therefore proceed as follows:
\begin{itemize}
    \item We work in units of the event horizon and rewrite Eq.~(\ref{eq:teukolsky2}) in order to remove the denominators
        \begin{equation}
        \begin{split}
        &  A(x)R''_s + B(x)R'_s + C(x) R_s=0\,,
        \end{split}
        \label{Teuk3}
        \end{equation}
    where the functions $A(x)$, $B(x)$, and $C(x)$ are given by the following:
        \begin{equation}\label{Teuk4}
        \begin{split}
        &  A(x)=f^2 (x+1)^2\,, \\
        &  B(x)= (s+1)f^2(x) (2(x+2)+(x+1)^2 f'/f)\,, \\
        &  C(x)=+(x+1)^2 \omega ^2 +2 i s (x+1) \omega f -i s (x+1)^2 \omega  f' \\
        & \;\;\;\;\;\;\;\;\;\;\;\, +s f \left ( (x+1)^2 f''+4 (x+1) f'+2 f-2 \right ) \\
        & \;\;\;\;\;\;\;\;\;\;\;\, -l (l+1) f +s (s+1) f\,,\nonumber
        \end{split}
        \end{equation}
    \item The lowest power term around $x=0$ of each coefficient can be written in the following form:
    \begin{equation}
        \begin{split}
        &  A(x)\sim x^2 \tau^2\,, \\
        &  B(x)\sim x (s+1) \tau^2\,, \\
        &  C(x) \sim \omega^2 - i \omega s \tau\,,\nonumber
        \end{split}
        \end{equation}
        where $\tau=\tau(\ell)$ depends on the choice of RBH.
    \item We then build the following characteristic equation:
        \begin{eqnarray}
          m(m-1) \tau^2  + m (s+1) \tau^2 + \omega (\omega - i s \tau)=0\,,
          \label{Eu-Cau}
        \end{eqnarray}
    whose solutions are the following:
    \begin{eqnarray}
          m_1=-s - \frac{i \omega}{\tau} \;\; ,\;\;   m_2=\frac{i \omega}{\tau}  
    \end{eqnarray}
\item It is then possible to conclude that Eq.~(\ref{eq:teukolsky2}) admits solutions near the singular point $x=0$ of the form given by Eq.~(\ref{eq:near}).
\end{itemize}
Explicitly, for the three RBHs in question, $\tau$ is given by the following:
        \begin{equation}
        \begin{split}
        & \tau_{\text{B}}=\frac{1-2 \ell^2}{\ell^2+1} \;\;,\\
        & \tau_{\text{H}}= (1-3  \ell^2)\;\;,\\
        & \tau_{\text{CGSV}}=1-\ell\;\;.\nonumber
        \end{split}
        \end{equation}
We notice that in the Schwarzschild limit $\ell \to 0$, all of the above reduce to $\tau=1$ as one could expect.

\section{PRBH time evolution}
\label{sec:appendixb}

Throughout the paper, in using $M$ to denote the masses of PRBHs, we never specified whether we were referring to the initial mass, the present mass of the PRBH undergoing evaporation, or another quantity. This question is relevant since the mass of an evaporating BH is of course a monotonically decreasing quantity. As we shall see, this question is intertwined with the question of what is the lifetimes of these PRBHs, and whether they have already evaporated by now: the aim of this Appendix is to address these questions.

In Sec.~\ref{subsec:spectra} we discussed the spectrum of photons emitted due to Hawking evaporation. In general, however, BHs will emit not only photons, but the whole spectrum of particles of the underlying theory. In fact, no specific coupling is required for this to occur. Enforcing energy conservation leads to the conclusion that the emission comes at the expense of the BH mass. Following the steps of Refs.~\cite{Page:1976df,Page:1976ki,Page:1977um,Chambers:1997ai,Taylor:1998dk}, and considering a field of spin $s$, the associated depletion function is defined as follows:
\begin{align}
{f_s}&=M^2 \int_0^{\infty}d\omega \frac{d^2N}{dtdE} \omega \nonumber \\
&= M^2\sum_{i,l,m}\frac{1}{2\pi} \int_0^{\infty}d\omega  \frac{n_i\Gamma^{s}_{l,m}(\omega)}{e^{\omega/T}\pm 1}\omega \nonumber \\
&=M^2\sum_{i,l}\frac{(2l+1)}{2\pi} \int_0^{\infty}d\omega  \frac{n_i\Gamma^{s}_{l}(\omega)}{e^{\omega/T}\pm 1}\omega\,,
\label{eq:depfunc}
\end{align}
where in the last equality we have exploited the fact that for spherically symmetric space-times the $(2l+1)$ different $m$ modes are degenerate, and $n_i$ is the number of degrees of freedom of the particle in question. It is worth noting that there is no reference to $\mu$, the mass of the field being evaporated by the BH. Given that the emission of particles with masses above the Hawking temperature $T$ is Boltzmann-suppressed, in practice working within the usually adopted approximation where the particles are considered massless for $\mu<T$, and absent at lower temperatures, is sufficiently accurate for our purposes (so we do not need to compute the GBFs for massive fields).

At a given particle temperature, summing over all the degrees of freedom accessible at a given BH temperature one can define the following function:
\begin{eqnarray}
\mathcal{F}=\mathcal{N}_0f_0+\mathcal{N}_{1/2}f_{1/2}+\mathcal{N}_1f_1+\mathcal{N}_{3/2}f_{3/2}+\mathcal{N}_2f_2\,, \nonumber \\
\label{eq:f}
\end{eqnarray}
where ${\cal N}_s$ count the degrees of freedom of spin $s$ such that $\mu<T$. This function is related to the BH mass loss rate as a function of time $t$ through the following differential equation:
\begin{equation}
\frac{dM}{dt}=-\frac{\mathcal{F}}{M^2}\,,
\label{eq:mass}
\end{equation}
which can easily be integrated numerically with initial condition given by $M(t=0)=M_0$. The lifetime of a BH is then determined by the time at which $M$ goes to $0$.

Let us begin by considering the case of evaporating Schwarzschild PBHs. Integrating Eq.~(\ref{eq:mass}) backwards, we compute the initial mass of a Schwarzschild PBH whose lifetime is roughly the age of the Universe ($t_U \sim 13.8\,{\text{Gyr}}$), finding $M_0 \sim 5 \times 10^{14}\,{\text{g}}$ in agreement with Ref.~\cite{Page:1976df}. On the other hand, we find the lifetime of a Schwarzschild PBH of initial mass $M_0=10^{15}\,{\text{g}}$ to be $\sim 10t_U$. However, of greater relevance to us is the question of how much does its mass vary from $M_0$ during the first $t_U$, i.e.\ from formation until today. We find that, during this time, the mass of the PBH is subject to a negligible (sub-\% level) decrease. In this light, and given that we have focused on PBHs with masses $M>10^{15}\,{\text{g}}$, which are far from having fully evaporated today, for our purposes we can safely approximate the BHs as being quasi-static throughout the lifetime of the Universe, while consequently denoting by $M$ the values of the PBH mass both at formation and today.

The above is valid for Schwarzschild PBHs. However, as we have already seen earlier, the three PRBHs we considered are all colder compared to their Schwarzschild counterparts (at a given mass), and the slight increase in the GBFs is not sufficient to compensate for these decreased temperatures. In other words, the considerations we drew in Sec.~\ref{subsec:spectra} lead us to conclude that these PRBHs have a longer lifetime compared to their Schwarzschild counterparts at a given mass, expectation that we have checked explicitly. For example, the lifetime of a Bardeen BH of $\ell=0.45r_H$ and initial mass $M_0=10^{15}\,{\text{g}}$ is $>10^3t_U$, and during the first $t_U$ the mass lost to evaporation is negligible. In short, the PRBHs we considered have lifetimes considerably longer than Schwarzschild PBHs of the same initial mass, and they lose a negligible fraction of mass throughout the age of the Universe: therefore, they are very far from having fully evaporated today, and the quasi-static approximation we have adopted is justified.

\bibliography{prbhi}

\end{document}